\documentclass[12pt]{article}

\usepackage{times}
\usepackage{fullpage}
\usepackage{amsfonts,amssymb,amsmath,amsthm}
\usepackage{latexsym}
\usepackage{gauss}
\usepackage{calc}
\usepackage{url}
\usepackage{thmtools}
\usepackage{thm-restate}
\usepackage{hyperref}
\usepackage[capitalise, nameinlink]{cleveref}
\usepackage{algorithm}
\usepackage{algorithmicx}
\usepackage{algpseudocode}
\usepackage[pdftex]{color}
\usepackage{graphicx}
\usepackage{enumitem}
\newcommand{\N}{\mathbb{N}}

\newcommand{\R}{\mathbb{R}}
\newcommand{\Z}{\mathbb{Z}}

\newcommand{\Acal}{\mathcal{A}}

\newcommand{\Pcal}{\mathcal{P}}
\newcommand{\Rcal}{\mathcal{R}}
\newcommand{\Mcal}{\mathcal{M}}
\newcommand{\Kcal}{\mathcal{K}}

\newcommand{\Scal}{\mathcal{S}}
\newcommand{\Tcal}{\mathcal{T}}

\newcommand{\Gcal}{\mathcal{G}}
\newcommand{\Xcal}{\mathcal{X}}

\newcommand{\atoms}{\mathrm{atoms}}

\newcommand{\CON}{\mathrm{CON}}

\newcommand{\Contract}{\mathrm{Contract}}

\newcommand{\rmin}{\mathrm{in}}
\newcommand{\rmout}{\mathrm{out}}

\DeclareMathOperator*{\argmin}{argmin}

\newcommand{\shore}{\mathrm{shore}}
\newcommand{\edges}{\mathrm{cutedges}}

\newcommand{\floor}[1]{\lfloor #1 \rfloor}
\newcommand{\ceil}[1]{\left\lceil #1 \right\rceil}

\def\01{\{0,1\}}

\newcommand{\ith}{i^{\scriptsize \mbox{{\rm th}}}}
\newcommand{\jth}{j^{\scriptsize \mbox{{\rm th}}}}

\newcommand{\ellth}{\ell^{\scriptsize \mbox{{\rm th}}}}

\newcommand{\tO}{\tilde O}
\newcommand{\Ecal}{\mathcal{E}}

\newtheorem{theorem}{Theorem}

\newtheorem{lemma}[theorem]{Lemma}

\newtheorem{proposition}[theorem]{Proposition}

\newtheorem{claim}[theorem]{Claim}

\theoremstyle{definition}

\newtheorem{definition}[theorem]{Definition}

\begin{document}
\title{Quantum complexity of minimum cut}
\author{Simon Apers\thanks{CWI, Amsterdam and ULB, Brussels. Email: smgapers@gmail.com} \and
Troy Lee\thanks{Centre for Quantum Software and Information, University of Technology Sydney. Email: troyjlee@gmail.com}
}
\date{}
\maketitle

\begin{abstract}
The minimum cut problem in an undirected and weighted graph $G$ is to find the minimum total weight of a set of edges whose removal disconnects $G$.
We completely characterize the quantum query and time complexity of the minimum cut problem in the adjacency matrix model.
If $G$ has $n$ vertices and edge weights at least $1$ and at most $\tau$, we give a quantum algorithm to solve the minimum cut problem using
$\tilde O(n^{3/2}\sqrt{\tau})$ queries and time.
Moreover, for every integer $1 \le \tau \le n$ we give an example of a graph $G$ with edge weights $1$ and $\tau$ such that 
solving the minimum cut problem on $G$ requires $\Omega(n^{3/2}\sqrt{\tau})$ queries to the adjacency matrix of $G$.  These results contrast with the 
classical randomized case where $\Omega(n^2)$ queries to the adjacency matrix are needed in the worst case even to decide if an unweighted graph is connected or not.

In the adjacency array model, when $G$ has $m$ edges the classical randomized complexity of the minimum cut problem is $\tilde \Theta(m)$.  
We show that the quantum query and time complexity are $\tilde O(\sqrt{mn\tau})$ and $\tilde O(\sqrt{mn\tau} + n^{3/2})$, respectively, where again the edge weights are between $1$ and $\tau$.
For dense graphs we give lower bounds on the quantum query complexity of $\Omega(n^{3/2})$ for $\tau > 1$ and $\Omega(\tau n)$ for any $1 \leq \tau \leq n$.

Our query algorithm uses a quantum algorithm for graph sparsification by Apers and de Wolf (FOCS 2020) and results on the structure of near-minimum cuts 
by Kawarabayashi and Thorup (STOC 2015) and Rubinstein, Schramm and Weinberg (ITCS 2018).
Our time efficient implementation builds on Karger's tree packing technique (STOC 1996).
\end{abstract}

\newpage


\section{Introduction}
Let $G = (V,w)$ be a weighted graph, where $w : \binom{V}{2} \rightarrow \R_{\ge 0}$ assigns a non-negative weight to every edge slot.  
We denote the edges of $G$, i.e.\ the edge slots that are given positive weight, by $E(G)$.  
For a nontrivial set $\emptyset \ne X \subsetneq V$ let 
$\Delta_G(X)$ be the set of edges of $G$ with exactly one endpoint in $X$ and 
one endpoint in $\overline{X} = V \setminus X$.  A \emph{cut} of $G$ is a set of 
edges of the form $\Delta_G(X)$ for some nontrivial set $X \subseteq V$.  We call $X$ and 
$\overline{X}$ the \emph{shores} of the cut.  The minimum cut problem is to determine the minimum of 
$\sum_{e \in \Delta_G(X)} w(e)$ over all non trivial subsets $X$.    This is equivalent to the minimum total weight of edges that need to be removed 
from $G$ in order to disconnect it.  We call this minimum value  $\lambda(G)$.  A set of edges $\Delta_G(X)$ realizing $\lambda(G)$ 
is called a \emph{minimum cut} of $G$.  If $G$ is unweighted $\lambda(G)$ is known as the \emph{edge connectivity} of $G$ and is the 
minimum number of edges whose removal disconnects $G$.

Computing the weight of a minimum cut of a graph is a 
fundamental computational problem that has been extensively studied in theoretical computer 
science since at least the 1960s \cite{GH61, FF62}.  It is also a problem of great practical importance, with applications 
to clustering algorithms \cite{Bot93} and evaluating network reliability, among others (see \cite{PQ82} for a survey of applications).  
Classically it is known that edge connectivity can be computed in nearly linear time even by 
\emph{deterministic} algorithms \cite{KT19,HRW20}.  For weighted graphs with $m$ edges, the weight of a minimum cut can be determined in 
nearly linear time\footnote{The $\tO(\cdot)$ notation hides polylogarithmic factors in its argument.} $\tO(m)$ by a randomized algorithm \cite{Karger00,MN20,GMW20} and in almost linear time $O(m^{1+o(1)})$ by a deterministic algorithm \cite{Li20}.

In this work we study quantum algorithms for the minimum cut problem
in two standard models for graph problems, the adjacency matrix and 
the adjacency array models.  In the adjacency matrix model a query consists of a pair $\{u,v\}$ of vertices, and the answer is $w(\{u,v\})$.  The adjacency array model allows $3$ types of queries: 
one can query the degree of a vertex $v$, the name of the $\ith$ neighbor of $v$, according to some arbitrary ordering, and the weight of the edge between 
$v$ and its $\ith$ neighbor.

For classical randomized algorithms, in the adjacency matrix model it is known that even deciding if a graph is connected or not requires $\Omega(n^2)$ queries in the worst case \cite{DHHM06}.
More recently, the randomized query complexity of edge connectivity was studied by Bishnu, Ghosh, Mishra and Paraashar \cite{BGMP20} in a common generalization of the adjacency matrix 
and adjacency array models called the \emph{local query} model.  This model allows queries to the degree of a vertex and to the $\ith$ neighbor of a vertex $v$, 
as in the adjacency array model, and also queries as to whether or not $\{u,v\}$ is an edge, as in the adjacency matrix model.  Over simple graphs $G$ with $m$ edges, they show an 
$\Omega(m)$ lower bound on the number of local queries needed by a randomized algorithm to succeed with probability $2/3$ for both the problems of determining the edge connectivity and 
outputting a cut realizing the edge connectivity \cite[Theorems 2 and 3]{BGMP20}.  

In this work we completely characterize the quantum query and time complexity of the minimum cut problem in the adjacency matrix model.  The complexity depends on what we call 
the \emph{edge-weight ratio}.  We say a graph has edge-weight ratio $\tau$ if the ratio of the largest weight of the graph to the smallest is at most $\tau$.
When the edge-weight ratio of an $n$-vertex graph is $\tau$, we give a bounded-error quantum algorithm to solve the minimum cut problem using $\tilde O(n^{3/2} \sqrt{\tau})$ queries and time 
in the adjacency matrix model (\cref{thm:time}).
For the unweighted case, i.e.\ the case $\tau = 1$, one can see this bound is tight as D\"{u}rr, Heiligman, H{\o}yer, and Mhalla \cite{DHHM06} show that even deciding if a graph is connected or not requires $\Omega(n^{3/2})$ quantum queries 
in the adjacency matrix model.  
We extend this bound by showing that for any $1 \le \tau \le n$ there is a graph family with edge-weight ratio $\tau$ for which solving the minimum cut problem requires $\Omega(n^{3/2} \sqrt{\tau})$ quantum queries to the adjacency matrix 
(\cref{thm:lower}).  For $\tau \ge n$ one can always use the trivial $O(n^2)$ algorithm, thus 
our results characterize the quantum query complexity of the minimum cut problem in the adjacency matrix model for any value of $\tau$.

For the adjacency array model, we give a bounded-error quantum algorithm that solves the minimum cut problem in an $n$ vertex, $m$ edge graph with edge-weight ratio $\tau$ using $\tO(\sqrt{mn\tau})$ quantum queries (\cref{thm:main}).
The quantum algorithm runs in time $\tO(\sqrt{mn\tau} + n^{3/2})$ (\cref{thm:time}).
In this case we do not know whether the bound is tight in all regimes.
For unweighted graphs ($\tau = 1$) the best lower bound we know of is $\Omega(n)$, which again follows from a lower bound for connectivity \cite{DHHM06}.
For any $\tau > 1$ we show that the minimum cut problem requires $\Omega(n^{3/2})$ quantum queries to the adjacency array (\cref{thm:adjlower2}).
Finally, for any $1 \leq \tau \leq 5n/8$ we show a lower bound of $\Omega(\tau n)$ on the number of quantum adjacency array queries for solving the minimum cut problem (\cref{thm:adjlowerL}).

In addition to computing the weight $\lambda(G)$ of a minimum cut, all of our upper and lower bounds also apply to outputting the edges or shores of a cut realizing $\lambda(G)$.

\subsection{Previous work}
We are not aware of any previous work on the quantum complexity of \emph{exact} global minimum cut.  The closest work to ours in topic is the recent paper of Apers and de Wolf \cite{AdW19}, which 
in particular shows that in a weighted graph a $(1+\varepsilon)$-approximation to the weight of a minimum cut can be found in time $\tilde O(n^{3/2}/\epsilon)$ in the adjacency matrix model and time 
$\tilde O(\sqrt{mn}/\epsilon)$ in the adjacency array model.  The sparsifier construction of Apers and de Wolf that yields this approximation also plays a key role in our algorithm.  

Another key work for us is the seminal paper of D\"{u}rr, Heiligman, H{\o}yer and Mhalla \cite{DHHM06} which gives tight bounds for the quantum complexity of many graph problems in both the adjacency matrix and adjacency array models.  In particular, 
they show that determining if a graph is connected or not, i.e.\ determining if the minimum cut value is zero or positive, requires $\Omega(n^{3/2})$ queries in the adjacency matrix 
model and $\Omega(n)$ queries in the adjacency array model.  These are still the best lower bounds we know of for simple graphs\footnote{We use the term simple graph to mean an undirected, unweighted 
graph with no self-loops and no multiple edges.} even for the more general problem of computing the edge connectivity. Indeed, 
we show the $\Omega(n^{3/2})$ connectivity lower bound in the adjacency matrix model is a tight lower bound even on the quantum complexity of edge connectivity.
In \cite{DHHM06} it is also shown that finding a spanning forest in the adjacency matrix model can be done with a quantum algorithm in queries and time $\tO(n^{3/2})$, which is a result we will make use of in our time efficient algorithm.

Two classical papers which inspired our algorithm are the works of Kawarabayashi and Thorup (KT) \cite{KT19} and Rubinstein, Schramm, and Weinberg (RSW) \cite{RSW18}.  
KT give the first near-linear time deterministic algorithm to compute the edge connectivity of a simple graph $G = (V,E)$.  A key idea of KT is to look at a \emph{contraction} of the original graph $G$.  Let 
$\Pcal = \{P_1, \ldots, P_k\}$ be a partition of $V$.  The contraction $G' = \Contract(G,\Pcal)$ is a multi-graph whose vertices are labeled by the sets in $\Pcal$ 
and which has all the edges of $G$ whose endpoints lie in different sets of $\Pcal$.  KT first check the cardinality of all \emph{star} cuts of the form $\Delta_G(\{v\})$, which can be 
done deterministically in linear time.  
To find the minimum non-star cut, KT show that any simple graph $G$ with minimum degree $d$ has a contraction $G'= \Contract(G,\Pcal)$ that 
preserves all of the near-minimum non-star cuts of $G$, but which has only $\tO(n/d)$ vertices and $\tO(n)$ edges.  Moreover, they 
show how to find such a contraction deterministically in near-linear time.  They then use Gabow's $\tilde O(\lambda(G)|E(G)|)$ mincut algorithm \cite{Gabow95} to find a minimum cut in $G'$.
If $G$ has $m$ edges then $\lambda(G') =\lambda(G) \leq m/n$, and as $|E(G')| \in \tilde O(n)$, this gives a time bound that is nearly linear in $m$.

RSW follow a similar high-level approach to give a classical randomized algorithm that computes the edge connectivity of a simple graph with
\emph{cut queries}.  In the cut query model, when the input is a graph $G$, an algorithm can query any nontrivial set $X$ and receive the answer $|\Delta_G(X)|$.  
RSW show that the edge connectivity of a simple graph can be computed with high probability by a randomized algorithm after $O(n \log(n)^3)$ cut queries.  
In fact, this algorithm finds \emph{all} minimum cuts of the graph.  The RSW algorithm again first evaluates all star cuts.
They then
remove the log factors from the KT result to show there is a partition $\Pcal$ of $V$ such that $G' = \Contract(G,\Pcal)$ preserves all 
near-minimum cuts of $G$ and has only $O(n)$ edges.\footnote{An $O(n)$ bound on the number of edges implies an $O(n/d)$ bound on the number of vertices in a black-box way.} 
Moreover, they show how to efficiently learn this contraction with cut queries.  
The log factors of the original KT proof were also removed via another algorithmic proof by Lo, Schmidt, and Thorup \cite{LST20}.  

Our quantum algorithm will follow the approach taken by RSW to learn such a contraction of $G$, as is detailed in the next section.

\subsection{Technical overview}
In this overview we focus on the adjacency matrix model.
Apart from the lower bound, most ideas carry over in a straightforward way to the adjacency array model.  
We start off by explaining the lower bound, as this clearly shows the origin of the~$n^{3/2} \sqrt{\tau}$ complexity.  

\paragraph*{Lower bound on the quantum query complexity.}
For the lower bound we construct a family of graphs on $2n$ vertices with edge weights in $\{1,\tau\}$.  Partition the $2n$ vertices into two sets $A$ and $B$ each of size $n$.  Make 
a complete graph among the vertices in $A$ where every edge has weight $\tau$ and do the same to $B$.  This ensures that $w(\Delta_G(X)) \ge \tau (n-1)$ for any $\emptyset \ne X \subset A$, 
and the same for $B$.  This large value gives us ``cover'' to hide either $k-1$ or $k$ edges of weight 1 between $A$ and $B$.  If $k < \tau(n-1)$ these edges will 
constitute the unique minimum cut, and thus an algorithm that outputs the weight of the minimum cut must determine if we hid $k-1$ or $k$ edges.  This is equivalent to 
determining if there are $k-1$ or $k$ marked items in a search space of size $n^2$, for which a quantum query lower bound of $\Omega(\sqrt{kn^2})$ is known \cite{NW99}.  In our case, with $k = \tau (n-1) -1$ 
this gives a bound of $\Omega(n^{3/2}\sqrt{\tau})$.  Thus we see that ultimately the lower bound for minimum cut boils down to the difficulty of counting for quantum algorithms.  
We will see how a similar task arises in the upper bound as well.  

\paragraph*{Upper bound on the quantum query complexity.}
We first describe a quantum algorithm for computing the edge connectivity of an unweighted graph.  We will follow the outline of the 
RSW cut query algorithm, which proceeds in the following way.  The algorithm first computes the degree of every vertex of $G$, thereby determining the minimum cardinality of a star cut.  The task is then reduced to finding 
the minimum cardinality of a non-star cut.  To do this, the RSW algorithm first produces an $\varepsilon$-\emph{cut sparsifier} of the graph, following an algorithm due to 
Bencz\'{u}r and Karger \cite{BK15}.  An $\varepsilon$-cut sparsifier of $G = (V,E)$ is a sparse weighted graph $H$ whose edge set is a subset of $E$, but where edges 
are allowed to be weighted.  For every nontrivial $X$ the weight of the cut $\Delta_H(X)$ in $H$ is within a factor of $1 \pm \varepsilon$ of $|\Delta_G(X)|$.  

For $\epsilon = 1/100$, the algorithm finds an $\varepsilon$-cut sparsifier $H$ of $G$.  The algorithm is able to write $H$ down in memory and then, without further queries, it can compute the weight of a minimum cut in $H$, say it is $\lambda(H)$, and enumerate all non-star cuts of $H$ whose weight is at most $(1+ 3 \epsilon) \lambda(H)$.  
With high probability this includes the shores of all non-star minimum cuts of $G$.  Let $\Tcal$ be the set of all shores of these cuts.  
The algorithm then computes the coarsest partition $\Pcal = \{P_1, \ldots, P_k\}$ of the vertex set with the property that for all $P_j \in \Pcal$ and $u,v \in P_j$ it holds 
that $u,v \in X$ or $u,v \in \overline{X}$ for all $X \in \Tcal$.  We call $\Pcal$ the set of atoms of $\Tcal$, denoted $\atoms(\Tcal)$. 
As $\Tcal$ is the set of shores of all non-star near-minimum cuts, this means that, for every $P_j \in \Pcal$, no non-star near-minimum cut has an edge with both endpoints in $P_j$;  
as $\Pcal$ is the coarsest partition with this property, among such partitions it minimizes the number of edges \emph{between} components of the partition.  A key 
fact is that $\Contract(G,\Pcal)$ is a sparse graph.

\begin{lemma}[\cite{KT19, RSW18,LST20}]
\label{lem:RSW}
Let $G = (V,E)$ be a simple $n$-vertex graph with minimum degree $d$.  For a nonnegative $\varepsilon < 1$, let 
$\Tcal = \{X : |X|, |\overline{X}| \ge 2 \mbox{ and } |\Delta_G(X)| \le \lambda(G) + \varepsilon d\}$, that is the set of shores of all non-star cuts whose weight is at most $\lambda(G) + \varepsilon d$, and let $G' = \Contract(G, \atoms(\Tcal))$.  Then 
$|E(G')| = O(n)$.
\end{lemma}
By the definition of $\Pcal$ in this lemma, one can also see that $G'$
preserves all of the non-star near-minimum cuts of $G$.  As we already know the minimum degree of $G$, to determine $\lambda(G)$ it suffices to compute 
the edge connectivity of $G'$.  For a query algorithm, to do this it suffices to learn the $O(n)$ edges of the graph $G'$; 
then one can compute the edge connectivity of $G'$ without further queries.  The edge connectivity of $G$ is then the minimum of the minimum degree of $G$ and the 
edge connectivity of $G'$.  

We phrase the RSW algorithm in an abstract way in terms of four computational primitives.  We indicate oracle access to $G$ by square brackets and put the 
parameters explicitly given to the routines in parentheses.
\begin{enumerate}
  \item FindMinStar$[G](\delta)$ --- a routine that given oracle access to $G$ finds the minimum weight of a star cut of $G$ with error probability at most $\delta$.
  \item Cut-Sparsifier$[G](\varepsilon, \delta)$ --- a routine that given oracle access to $G$ outputs an $\varepsilon$-cut sparsifer of $G$ with error probability at most $\delta$.
  \item LearnCutAtoms$(H, \lambda, \delta)$ --- a routine that given an explicit description of a graph $H$, a cut threshold $\lambda$, and an error probability $\delta$, 
  outputs $\Pcal$, the atoms of the shores of all cuts of weight at most $\lambda$, with error probability at most $\delta$.
  \item LearnContraction$[G](\Pcal, M, \delta)$ --- a routine that given oracle access to $G$ and a partition $\Pcal$ of the vertex set, learns $\Contract(G,\Pcal)$ if it has at most $M$ 
  edges and otherwise outputs NULL, again with error probability at most $\delta$.
\end{enumerate}
In \cref{thm:rsw_alg}, we show a general upper bound on the query complexity of edge connectivity in terms of the sum of the query complexity of the routines in steps (1), (2), and (4).  Step~(3) 
requires no queries.  It is somewhat surprising that a randomized algorithm designed for cut queries leads to an optimal quantum query algorithm in the adjacency matrix model.  We hope that phrasing 
the algorithm in this abstract way will make it easy to further apply it to other computational models.

In terms of quantum query complexity in the adjacency matrix model, the cost of the 4 steps are as follows.
Item~(1) can be done with $O(n^{3/2})$ queries by composing the $O(\sqrt{n})$ query quantum minimum finding algorithm over the $n$ vertices with the $n$ query classical algorithm to evaluate the degree of 
a vertex.  The quantum complexity of (2) was recently studied by Apers and de Wolf \cite{AdW19}.  They 
show that even an $\varepsilon$-\textit{spectral} sparsifier can be found in \emph{time} $\tilde O(n^{3/2}/\varepsilon)$ in the adjacency matrix 
model.  For our purposes, we take $\varepsilon=1/100$ giving an $\tO(n^{3/2})$ bound here. Item~(3) costs no queries as the routine is given an explicit description of $H$.  Item~(4) is very similar to the 
problem that we saw in the lower bound: we have to learn up to $M$ edges in a search space of size $O(n^2)$ which can be done with $O(n \sqrt{M})$ queries.  By \cref{lem:RSW}
we can take $M = O(n)$ resulting in an $O(n^{3/2})$ quantum query bound for this step.

These bounds when taken together imply a quantum algorithm for edge connectivity making $\tilde O(n^{3/2})$ queries in the adjacency matrix model.

\paragraph*{Extension to weighted graphs.}
The query complexity of steps~1--3 does not change for weighted graphs.  The complexity of step~4, however, depends on the upper bound $M$ on the number of 
edges in the graph $\Contract(G, \Pcal)$, which does depend on the edge weights.
To extend the above algorithm to weighted graphs, we prove the following generalization of \cref{lem:RSW}.
\begin{restatable}{lemma}{RSWweighted}
\label{thm:RSWweighted}
Let $G = (V,w)$ be a weighted graph with $|V| =n$ and where every edge has weight at most $\tau$.  Let $d = \min_{u \in V} w(\Delta_G(\{u\}))$.  For a nonnegative $\varepsilon < 1$, let 
$\Tcal = \{X : |X|, |\overline{X}| \ge 2 \mbox{ and } w(\Delta_G(X)) \le \lambda(G) + \varepsilon d\}$ and let $G' = \Contract(G, \atoms(\Tcal))$.  Then 
\[
w(E(G')) \le \frac{68 \tau n}{(1-\varepsilon)^2} \enspace .
\]
\end{restatable}
This lemma is tight as can be seen from the cycle graph with all edge weights $\tau$.  Because the bound necessarily depends on $\tau$, applying this lemma 
back to the cut query or sequential models does not seem to lead to good algorithms.\footnote{The randomized cut query complexity of minimum cut for weighted graphs was 
recently resolved using different techniques by Mukhopadhyay and Nanongkai \cite{MN20}.}
For quantum algorithms, however, it is exactly what is needed.

If the edge-weight ratio is $\tau$, for constant $\varepsilon$ \cref{thm:RSWweighted} implies an $O(\tau n)$ upper bound on the number of edges in the contracted graph $\Contract(G,\Pcal)$.
This means that the LearnContraction step can be performed with $O(n^{3/2}\sqrt{\tau})$ queries.  Together with the $\Omega(n^{3/2}\sqrt{\tau})$ query lower bound mentioned 
above we obtain the following tight characterization of 
the query complexity of minimum cut in the adjacency matrix model in terms of the edge-weight ratio.

\begin{restatable}{theorem}{thmquery} \label{thm:restate_main}
Let $G =(V,w)$ be an $n$-vertex weighted graph with edge-weight ratio $\tau$.  There is a quantum algorithm that finds the 
weight and shores of a minimum cut of $G$ with probability at least $3/4$ after 
$\tilde O(n^{3/2}\sqrt{\tau})$ queries to the adjacency matrix of $G$.  Moreover, there is a family of graphs with edge-weight ratio $\tau$ 
for which computing the weight of a minimum cut with bounded-error requires $\Omega(n^{3/2} \sqrt{\tau})$ quantum queries to the adjacency matrix.
\end{restatable}
The upper bound for this theorem is given in \cref{thm:main}, and the lower bound in \cref{thm:lower}.

\paragraph*{Upper bound on the quantum time complexity.}
Let us now consider the time complexity of the above algorithm, corresponding to the total number of queries and elementary gates in the quantum circuit model that the algorithm uses.  Steps~(1) and~(4) are ultimately applications of Grover's algorithm and 
can be implemented in time which is just a $O(\log(n))$ factor more than their query complexity.  For step~(2), Apers and de Wolf already give 
a time complexity upper bound of $\tO(n^{3/2}/\varepsilon)$.  Thus to get an upper bound on the time complexity it suffices to 
analyze the routine LearnCutAtoms$(H,\lambda,\delta)$ from step~(3).
Given a graph $H$, this subroutine requires us to output the atoms of $\Tcal$, where $\Tcal$ is the set of shores of all near-minimum cuts of $H$.  
For this discussion, one should take near-minimum cuts to mean cuts of weight at most $(1+1/100) \lambda(H)$.  
It is known that an $n$-vertex graph $H$ has at most $O(n^2)$ cuts of weight $< 3\lambda(H)/2$ \cite{HW96}.  Thus we know that $|\Tcal|$ is not too large.  
However, we still need to efficiently find these near-minimum cuts.  

\begin{figure}[htb]
\centering
\includegraphics[width=.5\textwidth]{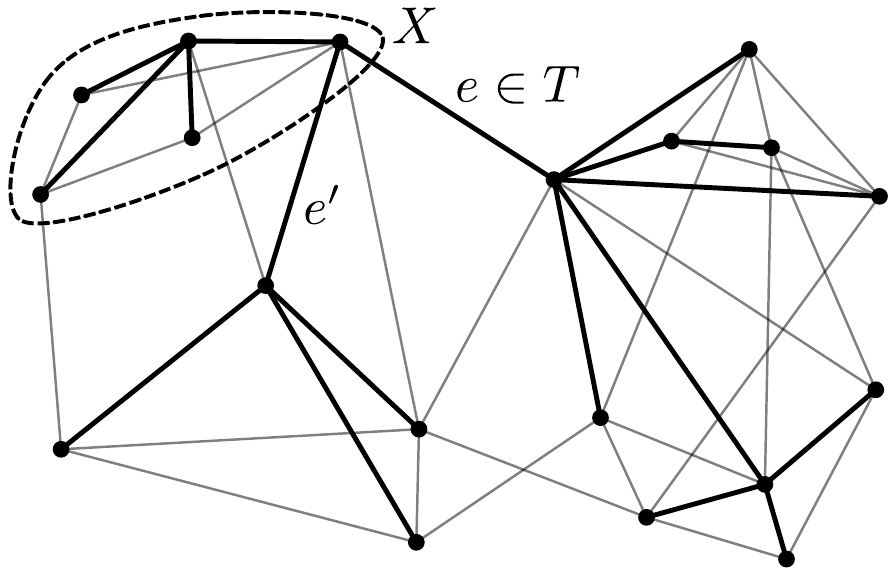}
\caption{Graph $H$ (thin grey edges) with spanning tree $T$ (thick black edges). The cut with shore $X$ 2-respects $T$ since $|\Delta_T(X)| = |\{e,e'\}| \le 2$.
There are at most $\binom{n}{2}$ such cuts.}
\label{fig:2-respecting}
\end{figure}

To do this we build on Karger's seminal work \cite{Karger00} that connects near-minimum cuts with tree packings.
Consider a spanning tree $T$ of $H$, as in \cref{fig:2-respecting}.
A cut in $H$ with shore $X$ is said to \emph{2-respect} $T$ if it cuts at most 2 edges of $T$, that is $|\Delta_T(X)| \leq 2$.
Karger showed how to efficiently construct a set of $O(\log n)$ spanning trees in $H$ so that every near-minimum cut 2-respects at least one of them.  As each 
tree has at most $n-1 + \binom{n-1}{2} = \binom{n}{2}$ 2-respecting cuts, this family of trees defines a set of shores $\Tcal'$ of cardinality $O(n^2 \log n)$ which necessarily contains $\Tcal$.
A graph can potentially contain $\binom{n}{2}$ minimum cuts, as witnessed by the cycle graph, thus this bound is nearly tight.  
Unfortunately, iterating over $\Tcal'$ is still too costly for us.

As we are only interested in $\atoms(\Tcal)$, and not $\Tcal$ itself, it suffices for us to find a set $\Scal$ such that $\atoms(\Scal) = \atoms(\Tcal)$.  We call such an $\Scal$ a \emph{generating set} for $\atoms(\Tcal)$.  Our next observation is that there necessarily exists a generating set for $\atoms(\Tcal)$ of size $O(n)$.  
This follows by a greedy argument: set $\Scal = \emptyset$ and iterate over all cut shores $X \in \Tcal$, adding $X$ to $\Scal$ iff $\atoms(\Scal \cup X) \ne \atoms(\Scal)$.
The resulting $\Scal$ has the same atoms as $\Tcal$.  Moreover, $|\Scal| \leq n-1$ since every element added to $\Scal$ creates at least one new atom, there are at most $n$ atoms in total, and $\Scal = \emptyset$ has 1 atom.
While a good start, this still leaves the problem of efficiently finding a small generating set.  

\begin{figure}[htb]
\centering
\includegraphics[width=.8\textwidth]{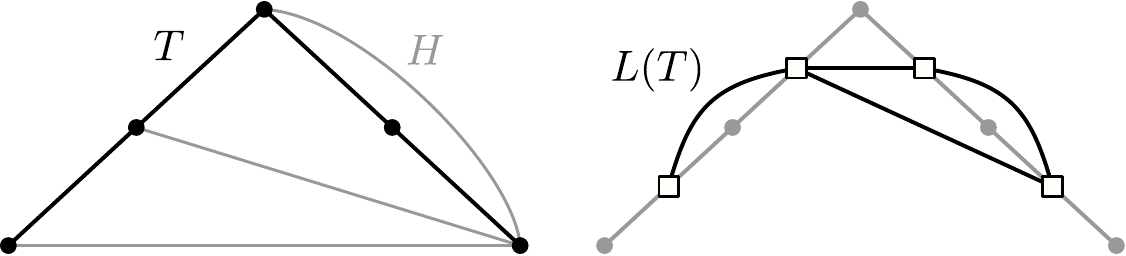}
\caption{Left: A spanning tree $T$ (thick black edges) of the graph $H$ (thin grey edges) with minimum cut $\lambda(H) = 2$. Right: The associated graph $L(T)$ with vertex set $E(T)$ and $f \in E(T)^{(2)}$ an edge of $L(T)$ iff $\shore(f)$ is the shore of a near-minimum cut in $H$ (in this case, a near-minimum cut is a cut of weight $\leq \frac{3}{2} \lambda(H)$).}
\label{fig:line-graph}
\end{figure}

We are able to give an explicit description of an $O(n \log(n))$ size generating set.
First consider a single spanning tree $T$ of $H$.
For any $f \in E(T) \cup E(T)^{(2)}$ we let $\shore(f)$ denote the cut shore such that $\Delta_T(\shore(f)) = f$. \footnote{As $\Delta_T(X) = \Delta_T(\overline{X})$, for uniqueness we define a root $r$ in $T$ and choose $\shore(f)$ so that it does not contain $r$.}
Now define an unweighted graph $L(T)$ whose vertex set is $E(T)$ and where $f \in E(T)^{(2)}$ is an edge of $L(T)$ iff $\shore(f)$ is a near-minimum cut of $H$ (i.e., $\shore(f) \in \Tcal$).
We show an example in \cref{fig:line-graph}.
Further, let $O(T) = \{e \in E(T): \exists X \in \Tcal: \Delta_T(X) = \{e\}\}$ index the set of near-minimum cuts that 1-respect $T$.
We prove the following lemma.
\begin{lemma}
Let $\Tcal' = \{X \in \Tcal : |\Delta_T(X)| \le 2\}$ be the shores in $\Tcal$ whose corresponding cuts $2$-respect $T$.  
If $F$ is a spanning forest of $L(T)$ then $\Scal(T) = \{\shore(f) \mid f \in E(F) \cup O(T)\}$ is a generating set for $\atoms(\Tcal')$.
\end{lemma}
Moreover, since $|E(F)| \le n-2$ and $|O(T)| \le n-1$ we have $|\Scal(T)| \le 2n-3$.  Taking the union of $\Scal(T)$ over all of the $\log(n)$ spanning trees $T$ of Karger's tree packing gives a generating set $\Scal$ for $\Tcal$ of size $O(n\log(n))$. 

We cannot explicitly write down the graph $L(T)$, but using an efficient data structure for evaluating 2-respecting cuts \cite{MN20,GMW20c} we can in $O(\log(n))$ time determine 
whether or not $\{e,e'\}$ is an edge of $L(T)$.
This essentially gives us adjacency matrix access to $L(T)$, and hence we can use the $\tO(n^{3/2})$ time quantum algorithm from~\cite{DHHM06} to construct a spanning forest $F$ of $L(T)$.
We note that it is conceivable that there exists an efficient classical algorithm to do this.
However this would require using further properties of $L(T)$ since classically computing a spanning forest in the adjacency matrix model requires $\Omega(n^2)$ queries.

Once we have the $O(n \log n)$ size generating set $\Scal$, we still cannot naively compute the atoms of $\Scal$ because this would again be too costly.
Rather, we find the atoms of $\Scal$ in $\tO(n)$ time by combining a random hashing scheme with an efficient data structure based on Euler tour trees \cite{HK95}.
This shows that a quantum algorithm can implement step~(3), LearnCutAtoms, in time $\tO(n^{3/2})$.  Note that this running time is independent of the kind of oracle access we have to $G$.  
This gives the following theorem.

\begin{restatable}{theorem}{thmtime} \label{thm:time}
Let $G =(V,w)$ be an $n$-vertex weighted graph with $m$ edges and edge-weight ratio $\tau$.  There is a quantum algorithm that finds the weight and shores of a minimum cut of $G$ with probability at least 
$2/3$ in query and time complexity $\tO(n^{3/2}\sqrt{\tau})$ in the adjacency matrix model and $\tO(\sqrt{mn\tau} + n^{3/2})$ in the adjacency array model.
\end{restatable}

\subsection{Open problems}
A few open problems remain from this work.
\begin{enumerate}
  \item
  In the adjacency array model there remains a significant gap between the upper and lower bounds we are able to show.
  For dense graphs the upper bound is $\tO(n^{3/2} \sqrt{\tau})$ and we have the lower bounds $\Omega(n^{3/2})$ for $\tau > 1$ and $\Omega(\tau n)$ for $1 \leq \tau \leq n$.
  We suspect that the quantum query complexity of the minimum cut problem in the adjacency array model is $\widetilde\Theta(n)$ for simple graphs ($\tau = 1$) and $\widetilde\Theta(\sqrt{mn\tau})$ for weighted graphs ($1 < \tau \leq m/n$), 
  but were unable to prove this.
  \item
  We have given a quantum algorithm with running time $\tO(m + n^{3/2})$ for the subroutine LearnCutAtoms.
  By building on our insights we believe that this routine can even be performed by a classical randomized algorithm in near-linear time $\tO(m)$.
  This would improve the running time of our quantum algorithm for the minimum cut problem in the adjacency array model from 
  $\tO(\sqrt{mn\tau} + n^{3/2})$ to $\tO(\sqrt{mn\tau})$.
  It also seems of more general interest, giving a weighted (but potentially randomized) generalization of the algorithm by Kawarabayashi and Thorup \cite{KT19} for finding a contraction of $G$ that 
  preserves all near-minimum cuts and only has $O(\tau n)$ total weight of edges.
  \item
  What is the quantum complexity of determining a $(1+ \varepsilon)$-approximation of the minimum cut weight?
  Apers and de Wolf \cite{AdW19} gave a $(1+\varepsilon)$-approximation algorithm with time and query complexity $\tO(\sqrt{mn}/\varepsilon)$ in the adjacency array model.
  For the unweighted case, our algorithm improves this in terms of query complexity by exactly computing the minimum cut with $\tO(\sqrt{mn})$ queries.
  Can one approximate the weight of a minimum cut in an unweighted graph with even fewer queries?
\end{enumerate}


\section{Preliminaries}
For a natural number $n \ge 1$ we let $[n] = \{1, \ldots, n\}$.  For a real number $x$ we let $\lfloor x \rceil$ denote the closest integer to $x$.

\subsection{Graph basics and notation}
Let $V$ be a finite set and $V^{(2)}$ the set of all subsets of $V$ of cardinality $2$.  We represent a weighted undirected 
graph as a pair $G = (V,w)$ where $w: V^{(2)} \rightarrow \R$ is a non-negative function.  We let $V(G)$ be the vertex set of a graph $G$ and 
$E(G) = \{ e \in V^{(2)}: w(e) > 0\}$ be the set of edges of $G$.  We extend the weight function to sets $S \subseteq V^{(2)}$ by $w(S) = \sum_{e \in S} w(e)$.
We say that $G$ is \emph{simple} if $w: V^{(2)} \rightarrow \{0,1\}$ and in this case also denote $G$ as $G = (V,E)$, where $E$ is the set of edges.  We call the 
ratio of the largest edge weight of $G$ to the smallest the \emph{edge-weight ratio} of $G$.

For a subset $X \subseteq V$ we use the shorthand $\overline{X} = V \setminus X$, and we say $X$ is \emph{non-trivial} if $\emptyset \ne X \subsetneq V$.  
For disjoint sets $X,Y \subseteq V$ we use $E(X,Y)$ for the set of edges with one endpoint in $X$ and one endpoint in $Y$.
For a non-trivial set $X$, let $\Delta_G(X) = \{ \{i,j\} \in E(G) : i \in X, j \in \overline{X}\}$ be the set of edges 
of $G$ with one endpoint in $X$ and one endpoint in $\overline{X}$.  A \emph{cut} of $G$ is a set of the form $\Delta_G(X)$ for some non-trivial 
set $X$.  We call $X$ and $\overline{X}$ the \emph{shores} of the cut $\Delta_G(X)$.  We call a cut of the form $\Delta_G(\{u\})$ a \emph{star} cut, and refer 
to all other cuts as \emph{non-star} cuts.  The \emph{weight} of a cut $S$ is $w(S)$, which in the 
case of a simple graph equals $|S|$.  We let $\lambda(G) = \min_{\emptyset \ne X \subsetneq V} w(\Delta_G(X))$ be the minimum weight of a cut in $G$.  
We call a cut realizing this bound a \emph{minimum cut}.  We call a cut $\Delta_G(X)$ satisfying $w(\Delta_G(X)) \le \alpha \lambda(G)$ an $\alpha$-near minimum cut. 
In the case where $G$ is simple we call $\lambda(G)$ the \emph{edge connectivity} of $G$.  
We will only use the term edge connectivity in the context of unweighted graphs.

\begin{definition}[Vertex Contraction]
Let $G = (V,w)$ be a weighted graph and $\Pcal = \{S_1, \ldots, S_k\}$ be a partition of $V$.  Define $\Contract(G,\Pcal)$ to be 
the $k$-vertex weighted graph $G' = (\Pcal, w')$ where $w'(\{S_i,S_j\}) = w(E(S_i,S_j))$
for each $\{S_i, S_j\} \in \Pcal^{(2)}$.
\end{definition}
\noindent
Note that as long as $|\Pcal| \ge 2$ it will hold that $\lambda(\Contract(G,\Pcal)) \ge \lambda(G)$.  

We will also need to make use of graph sparsifiers. 
\begin{definition}[Cut sparsifier]
 For a weighted graph $G = (V,w)$ and $\varepsilon > 0$ an $\varepsilon$-\emph{cut sparsifier} $H = (V,w')$ of $G$ 
satisfies 
\begin{enumerate}
\item $H$ is a reweighted subgraph of $G$, that is $w'(e) > 0$ only if $w(e) > 0$.
\item It holds that $(1-\varepsilon) w(\Delta_G(X)) \le w'(\Delta_H(X)) \le (1+\varepsilon) w(\Delta_G(X))$ for all $\emptyset \neq X \subsetneq V$.
\end{enumerate}
\end{definition}
Cut sparsifiers were first defined by Bencz\'{u}r and Karger \cite{BK15} who showed that a weighted graph $G$ has an $\varepsilon$-cut sparsifier $H$ 
with $O(n \log(n)/\epsilon^2)$ edges, and $H$ can be constructed by a randomized algorithm in time $O(m \log^3(n))$.  
Fung, Hariharan, Harvey and Panigrahi \cite{FHHP19} have since shown that a cut sparsifier with the same bound on the number of edges can be constructed 
by a randomized algorithm in time $O(m) + \tilde O(n/\varepsilon^2)$, and Batson, Spielman and Srivastava \cite{BSS12} have given a deterministic 
polynomial time construction of sparsifiers with only $O(n/\varepsilon^2)$ edges.

\subsection{Atoms}
A family of subsets $\Tcal = \{X_1,\ldots,X_k\}$ of $V$ induces a partition of $V$ given by the regions in the Venn diagram of $\Tcal$.
We call the resulting sets of this partition the \emph{atoms} of $\Tcal$:
\begin{definition}[Atoms]
Let $V$ be a finite set and let $\Tcal = \{X_1, \ldots, X_k\}$ where each $X_i  \subseteq V$.  Define $\atoms(\Tcal) = \{A_1, \ldots, A_\ell\}$ 
to be a partition of $V$ such that 
\begin{enumerate}
  \item For any $A_j \in \atoms(\Tcal)$ and $u,v \in A_j$ it holds that for all $X_i \in \Tcal$ either $u,v \in X_i$ or 
  $u,v \in \overline{X}_i$.
  \item $\atoms(\Tcal)$ is the coarsest partition with property~(1).
\end{enumerate}
\end{definition} 

\begin{definition}[Generating set]
Let $V$ be a finite set and $\Tcal$ a set of subsets of $V$.  We say that $\Scal \subseteq \Tcal$ is a \emph{generating set} for 
$\atoms(\Tcal)$ if $\atoms(\Scal) = \atoms(\Tcal)$.
\end{definition}

\begin{proposition}
\label{prop:atom_cup}
Let $V$ be a finite set and $\Tcal_1, \Tcal_2$ two sets whose elements are subsets of $V$.  Let $\Scal_1, \Scal_2$ be generating sets 
for $\atoms(\Tcal_1), \atoms(\Tcal_2)$ respectively.  Then $\Scal_1 \cup \Scal_2$ is a generating set for $\atoms(\Tcal_1 \cup \Tcal_2)$.
\end{proposition}

\begin{proof}
As $\Scal_1 \subseteq \Tcal_1, \Scal_2 \subseteq \Tcal_2$ by the definition of a generating set,
$\Scal_1 \cup \Scal_2 \subseteq \Tcal_1 \cup \Tcal_2$ and $\atoms(\Tcal_1 \cup \Tcal_2)$ is a refinement of $\atoms(\Scal_1 \cup \Scal_2)$.  
Now we show that for any $u,v$ that are in different sets of $\atoms(\Tcal_1 \cup \Tcal_2)$ there is a set $S \in \Scal_1 \cup \Scal_2$ which 
separates them.  This will imply that in fact $\atoms(\Tcal_1 \cup \Tcal_2) = \atoms(\Scal_1 \cup \Scal_2)$.

If $u,v$ are in different sets of $\atoms(\Tcal_1 \cup \Tcal_2)$ then there must be a $T \in \Tcal_1 \cup \Tcal_2$ which separates them.  Suppose 
without loss of generality that $T \in \Tcal_1$.  Then since $\atoms(\Scal_1) = \atoms(\Tcal_1)$ and $u,v$ are in different sets of $\atoms(\Tcal_1)$, 
there must be an $S \in \Scal_1$ which separates $u$ and $v$.  This completes the proof.
\end{proof}

\subsection{Quantum query and computational models}
For general background on the quantum query model we refer the reader to \cite{HLS07}. 
Here we restrict ourselves to describing the quantum 
implementation of the input oracles in the adjacency matrix and adjacency array models.

In the adjacency matrix model, on input a weighted graph $G = (V,w)$, classically one can query any $\{u,v\} \in V^{(2)}$ and receive the answer 
$w(\{u,v\})$.  We now describe how to model this by a quantum query.  We will assume that the 
edge weights are given as binary decimal numbers with $M_1$ bits before the decimal and $M_2$ bits after the decimal for a total of $M = M_1 + M_2$ bits.  
The state of the quantum query 
algorithm will have three registers, a query register, an answer register, and a workspace register.  The 
state of the algorithm will in general be in a superposition of the basis states $|\{u,v\} \rangle |b \rangle | a \rangle$ where $\{u,v\} \in V^{(2)}, b \in \{0,1\}^M$ 
and $a \in \Acal$ for an arbitrary finite set $\Acal$.  
On input graph $G = (V,w)$, the input oracle $\mathsf{O}_G$ acts on a basis state $|\{u,v\} \rangle |b \rangle | a \rangle$ as
\[
\mathsf{O}_G |\{u,v\} \rangle |b \rangle | a \rangle = |\{u,v\} \rangle |b \oplus w(\{u,v\}) \rangle | a \rangle \enspace.
\]
 
In the adjacency array model, on input a weighted $n$-vertex graph $G = (V,w)$ one can make two types of queries.  In the first type, one can query a vertex $v \in V$ and receive its degree $\deg(v)$.
The second type is specified by a family of functions $\{f_v : [\deg(v)] \rightarrow V\}_{v \in V}$ such that $f_v(i)$ corresponds to the $\ith$ neighbor of vertex $v$ (according to some arbitrary but fixed ordering).  
A query consists of a pair $(v,i)$ for $i \in [\deg(v)]$ and the returned answer is the pair
$(f_v(i), w(\{v,f_v(i)\}))$.  In this paper we will only 
need to model the second type of query quantumly.  This is because our upper bounds are larger than $n$ so we can let the algorithm classically 
query all degrees at the start of the algorithm, and in our lower bound on the query complexity of edge connectivity for weighted graphs we assume the 
algorithm already knows the degree of every vertex.  
The state of the quantum query algorithm will again have a query register, an answer register, and a workspace register, with the state of the algorithm 
in general being in a superposition of the basis states $|(v,i)\rangle |x \rangle |b \rangle | a \rangle$ where $v \in V, i \in [\deg(v)], x \in \{0,\ldots, n-1\}, b \in \{0,1\}^M$, 
and $a \in \Acal$ for an arbitrary finite set $\Acal$.  We further let $\tau : V \rightarrow \{0,1,\ldots, n-1\}$ be a bijection where $|V| = n$.  Then the input oracle $\mathsf{O}_G$ acts on a basis state in the following way:
\[
\mathsf{O}_G |(v,i)\rangle |x \rangle |b \rangle | a \rangle = |(v,i) \rangle |x + \tau(f_v(i)) \bmod n \rangle |b \oplus w(\{v,f_v(i)\}) \rangle | a \rangle \enspace.
\]

In \cref{sec:time-efficient} we will further show that our query algorithms can be implemented in a time efficient manner.  We analyze the time complexity in terms of the standard quantum circuit model augmented with two types of oracles.  
One is the oracle for the input, either in the adjacency matrix or array model, and the second is an oracle to a classical memory of $\tO(n)$ bits.  The latter corresponds to a \emph{quantum random-access-memory} or \emph{QRAM}.  We further assume that 
we can classically update a value in this $\tO(n)$ bit classical memory in time $\tO(1)$.  
The assumption of QRAM access is also required for the time efficiency of the sparsifier construction in \cite{AdW19} which our algorithms build on, and in fact is a necessary (but sometimes inexplicit) assumption in the time analysis of many quantum algorithms for graph problems, e.g.~\cite{DHHM06,Amb06,B13}.

\subsection{Quantum algorithmic primitives}
We now go over the quantum subroutines we will need.  We need several variants of quantum search.
\begin{theorem}[Quantum search \cite{Gro97}]
\label{thm:grover}
Given oracle access to a string $x \in \{0,1\}^N$ such that $|x| > 0$,
there is a quantum algorithm that with probability at least $9/10$ returns an $i$ such that $x_i = 1$.
The algorithm makes $O(\sqrt{N})$ queries to $x$ and has time complexity $O(\sqrt{N} \log(N))$.
\end{theorem}

\begin{theorem}[{Exact quantum search, \cite[Theorem 4]{BHMT02}}]
\label{thm:qexact}
Given a positive integer $k$ and oracle access to a string $x \in \{0,1\}^N$ with $|x| = k$,
there is a quantum algorithm that returns an $i$ such that $x_i=1$ with certainty.
The algorithm makes $O(\sqrt{N/k})$ queries to $x$ and has time complexity $O(\sqrt{N/k} \log(N))$.
\end{theorem}

\begin{theorem}[{Based on \cite[Theorem 3]{BCWZ99}}]
\label{thm:qsearch}
Given $t,N \in \N$ with $1 \le t \le N$ and oracle access to $x \in \{0,1\}^N$,  there is a quantum algorithm such that
\begin{itemize}
  \item if $|x| \le t$ then the algorithm outputs $x$ with certainty, and
  \item if $|x| > t$ then the algorithm reports so with probability at least $9/10$.
\end{itemize}
The algorithm makes $O(\sqrt{tN})$ queries to $x$ and has time complexity $O(\sqrt{tN} \log(N))$.
\end{theorem}

\begin{proof}
Initialize $S = \emptyset$.  For $k = t$ down to $1$, do: (i) run exact quantum search (from \cref{thm:qexact}) on $x$ with parameter $k$, returning an index $i$, (ii) query $x_i$ and if $x_i = 1$ then 
add $i$ to $S$ and ``unmark'' $x_i$ for all future iterations, i.e. implicitly return $x_i = 0$ to future queries of the algorithm.

Finally, run normal quantum search (from \cref{thm:grover}) on the indices of $x$ outside of $S$ to check that there are no more solutions.
If this returns an $i \not \in S$ such that $x_i = 1$, then report $|x| > t$, otherwise return the string $y$ where $y_i =1$ if $i \in S$ and 
$y_i = 0$ otherwise.

The query complexity of the algorithm is
\[
O\left( \sum_{k=1}^{t} \sqrt{\frac{N}{k}} \right) + O(\sqrt{N}) = O(\sqrt{t N}) \enspace ,
\]
and its time complexity is similarly $O(\sqrt{tN} \log(N))$, as claimed.

For correctness, first note that if $|x| > t$ then necessarily an index $i$ such that $x_i = 1$ is remaining in the final step.
Quantum search \cref{thm:grover} will find such an index with probability at least $9/10$.
It remains to prove that $x$ is learned with certainty if $|x| \le t$.
To this end, assume for contradiction that $|S| < |x|$.
Then necessarily there was an iteration $k'$ between $t$ and $1$ such that $k' = |x|$.
In such case, however, the remaining $k'$ runs of exact quantum search will each return a nonzero index, and so all nonzero indices will be found.
This proves that necessarily all indices are found in the first $t$ iterations of exact quantum search, and hence the final quantum search step cannot find an additional nonzero index.
\end{proof}

\begin{theorem}[Quantum minimum finding \cite{DH96}]
\label{thm:qmin}
Let $N,M \in \N$ be positive integer and $f: [N] \rightarrow \R$.  There is a quantum algorithm that with probability at least $2/3$ outputs an element of $\argmin_{i \in [N]} f(i)$.
The algorithm makes $O(\sqrt{N})$ oracle calls to $f$ and has time complexity $\tO(\sqrt{N})$.
\end{theorem}

\begin{theorem}[{\cite[Theorem 1]{AdW19}}]
\label{thm:qsparsifier}
Let $G$ be a weighted $n$-vertex graph with $m$ edges.  There is a quantum algorithm that with high probability outputs an explicit description of an 
$\varepsilon$-cut sparsifier $H$ of $G$ with $\tilde O(n/\varepsilon^2)$ edges in query and time complexity $\tilde O(\sqrt{mn}/\varepsilon)$ in the adjacency array model or $\tilde O(n^{3/2}/\epsilon)$ in the adjacency matrix model.
\end{theorem}
Apers and de Wolf actually show a stronger theorem than this in that their algorithm can output a spectral sparsifier instead of 
just a cut sparsifier.  We will not need this additional property, however.

\subsection{Problems related to minimum cuts}
Let $G=(V,w)$ be a weighted graph.  There are three outputs related to a minimum cut of $G$ that one could want 
from an algorithm: the weight of a minimum cut, the shores of a minimum cut, or the edges in a minimum cut.
The relationship between the complexity of these problems is not always obvious, and can depend on the computational 
model one is studying.  All the upper and lower bounds we prove in this paper apply to all three problems.

Say the edge-weight ratio of $G$ is $\tau$.
As an example of how we can apply the quantum search algorithm \cref{thm:qsearch}, 
we show that, given the shores of a minimum cut in $G$, a quantum algorithm can also find the edges of the cut with 
$O(n^{3/2} \sqrt{\tau})$ and $O(\sqrt{mn \tau})$ queries in the adjacency matrix and array models respectively.  As this matches the complexity of 
our upper bounds, we will only explicitly mention finding the weight and shores of a minimum cut
in \cref{thm:main}.  

\begin{proposition}
Let $G = (V,w)$ be an $n$-vertex weighted graph with edge-weight ratio $\tau$.   Let $\Delta_G(X)$ be a minimum cut of $G$.  
Given $X$, a quantum algorithm can with probability at least $3/4$ output $\Delta_G(X)$ with $O(n^{3/2}\sqrt{\tau})$ queries and time complexity $\tO(n^{3/2}\sqrt{\tau})$ in the adjacency matrix model, and $O(\sqrt{mn\tau})$ queries and time complexity $\tO(\sqrt{mn\tau})$ in the adjacency array model.
\end{proposition}

\begin{proof}
Consider the adjacency matrix model first.  With $O(n)$ queries and time $O(n \log(n))$ we can identify the smallest and largest edge weights of $G$ except error probability at most $1/8$.  Thus by rescaling we will 
henceforth assume that the smallest edge weight is $1$ and largest edge weight is at most $\tau$.

Let $x \in \{0,1\}^{\binom{n}{2}}$ denote a bit string labeled by elements of $V^{(2)}$ and set $x(\{u,v\}) = 1$ iff $\{u,v\} \in E(G)$ and $u$ and $v$ are not both in 
$X$ or both in $\overline{X}$.  Given $X$, a query to $x$ can be answered by a single query to the adjacency matrix of $G$.  As the largest weight of an edge of $G$ is at most $\tau$ and $\Delta_G(X)$ is a minimum cut, 
$w(\Delta_G(X)) \le \tau(n-1)$.  As every edge of $G$ has weight at least $1$ we also have $|x| \le \tau (n-1)$.  Thus by \cref{thm:qsearch}, except with error probability $1/8$, we can learn $x$, and therefore also $\Delta_G(X)$, with 
$O(n^{3/2}\sqrt{\tau})$ queries and time $\tO(n^{3/2}\sqrt{\tau})$.

The statement for the adjacency array model follows from \cref{thm:qsearch} by a similar argument. 
\end{proof}


\section{Number of edges in near-minimum cuts}
In this section, we generalize \cref{lem:RSW} to weighted graphs.  Our proof follows that of Rubinstein, Schramm, and Weinberg \cite{RSW18}.

\RSWweighted*
Before proving this lemma we first state and prove a claim.
\begin{claim}
\label{clm:sizeK}
Let $V$ be a finite set of cardinality $n$ and $r \le n$ be a positive integer.  Let $\Tcal = \{X_1, \ldots, X_k\}$ where each $X_i \subseteq V$.  Let 
$\Tcal_0 = V$ and for $i = 1, \ldots, k$ let $\Tcal_i = \{X_1, \ldots, X_i\}$.  Suppose that $\Tcal$ has the property that for all $i = 0, \ldots, k-1$ 
there is a set $A_j \in \atoms(\Tcal_i)$ that is refined into two sets each of cardinality $\ge r$ in $\atoms(\Tcal_{i+1})$.  Then $|\Tcal| \le \frac{n}{r}-1$.
\end{claim}

\begin{proof}
To each $\Tcal_i$ for $i = 1, \ldots, k$ we associate a binary tree $B_i$.  Each vertex of $B_i$ has a label, which will be an element of $\cup_{j=0}^i \atoms(\Tcal_j)$.  
The tree $B_1$ has root $v$, labeled by $V$, and two children $v_0, v_1$ labeled by the two elements $X_1, \overline{X}_1 \in \atoms(\Tcal_1)$.  Note that by definition 
$|X_1|, |\overline{X}_1| \ge r$.  

In general, the tree $B_{i+1}$ is formed from $B_i$ as follows.  Initially, set $B_{i+1}=B_i$.  Then for every leaf $u$ of $B_i$ which is labeled by a set $Y \in \atoms(\Tcal_i)$ of size $\ge 2r$, 
if $Y$ is refined into sets $Y_1, Y_2$ in $\atoms(\Tcal_{i+1})$, then in $B_{i+1}$ the node $u$ is given two children labeled by $Y_1$ and $Y_2$, respectively.  Note that this construction 
has the property that only internal vertices of $B_i$ that are labeled by sets of size $\ge 2r$ have children.  Call a vertex \emph{big} if it is labeled by a set of size $\ge r$ and \emph{small} 
otherwise.  By construction, every internal vertex of $B_i$ has at least one big child.

Let $b_i$ be the number of big leaves in $B_i$.  We now show by induction that $i \le b_i - 1$.  This will prove the claim as the leaves of $B_i$ partition $V$ and therefore 
$b_i \le n/r$.  

For $i=1$ we have that $b_i = 2$ since $|X_1|, |\overline{X}_1| \ge r$, thus the base case holds.  Now suppose that $i \le b_i-1$, we will show that $i+1 \le b_{i+1}-1$.  
By definition of $\Tcal$, there must be some set $Y \in \atoms(\Tcal_i)$ which is refined into two sets $Y_1, Y_2$ both of cardinality at least $r$ in $\atoms(\Tcal_{i+1})$.  
Further, $Y$ will label some leaf of $u$ of $B_i$ and $u$ will have two children which are big in $B_{i+1}$.  Any other big leaf of $B_i$ which becomes an internal vertex 
of $B_{i+1}$ must have at least one child which is big.  This shows that $b_{i+1} \ge b_i + 1$ and gives the inductive step.
\end{proof}

Now we are ready for the proof of \cref{thm:RSWweighted}.
\begin{proof}[Proof of \cref{thm:RSWweighted}]
Let $\alpha=\beta = \frac{1}{4} (1-\varepsilon)$ so that $\alpha + \beta \le \frac{1}{2}(1-\varepsilon)$.  
Let $K \subseteq \Tcal$ be formed as follows.  Initialize $K$ to be empty.  Then do the following: while there is an $X \in \Tcal$ such that 
there is an $A \in \atoms(K), A_1, A_2 \in \atoms(K \cup X)$ such that $A = A_1 \cup A_2$ and $|A_1|, |A_2| \ge \frac{\beta d}{\tau}$,
add $X$ to $K$.  
By \cref{clm:sizeK}, at the end of this process $|K| \le \frac{\tau n}{\beta d}$.  
Let $\Kcal = \cup_{X \in K} \Delta_G(X)$ be the set of edges of cuts with shores in $K$.  Throughout this proof, cuts will 
always be with respect to $G$ and we will henceforth drop the subscript to simply write $\Delta(X)$.  

Let $S \subseteq V$ be the set of vertices $v$ such that $w(E(v, V\setminus \{v\}) \cap \Kcal)  \ge \alpha \cdot w(v)$.  
We say that $v \in V$ is \emph{small} if for the $A \in \atoms(K)$ with $v \in A$ there is an $X \in \Tcal$
such that $\atoms(K \cup X)$ refines $A$ into $A_1, A_2$ with $v \in A_1$ and $|A_1| < \frac{\beta d}{\tau}$.  

\begin{claim}
\label{clm:small}
If $v$ is small then $v \in S$.
\end{claim}

\begin{proof}
Let $X \in \Tcal$ be the shore of a cut which witnesses that $v$ is small.  Let us assume without loss of generality that $v \in X$.  
Suppose for contradiction that $v \not \in S$.  There are three possibilities for an edge $\{u,v\}$:  either $\{u,v\} \in \Kcal$, 
or $u \in A_1$, or $u \in A_2$.  Let the total weight of these kind of edges be $w_\Kcal, w_1, w_2$, respectively.  
Thus $w(v) = w_\Kcal + w_1 + w_2$.  We further know that $w_\Kcal < \alpha w(v)$ by the assumption that $v \not \in S$ and that $w_1 < \beta d$ since $|A_1| < \frac{\beta d}{\tau}$ 
and the maximum edge weight is $\tau$.  This means $w_2 > w(v) - \alpha w(v) - \beta d$.  Further note that $v$ contributes weight at least 
$w_2$ to the weight of $\Delta(X)$.  

As $\Delta(X)$ is not a star cut, we can consider the cut $\Delta(X')$ where $X' = X \setminus \{v\}$.  We claim that $w(\Delta(X')) < \lambda$, which is 
a contradiction.  The only difference between $w(\Delta(X))$ and $w(\Delta(X'))$ is the contribution of $v$.  The weight of edges involving $v$ in $\Delta(X')$ 
is at most $w_\Kcal + w_1 < \alpha w(v) + \beta d$.  Thus 
\begin{align*}
w(\Delta(X)) - w(\Delta(X')) &\ge w_2 - (w_\Kcal + w_1) \\
&> w(v) - 2\alpha w(v) - 2 \beta d \\
&\ge d(1-2\alpha - 2\beta) \\
&\ge \varepsilon d \enspace ,
\end{align*}
implying that $w(\Delta(X')) < \lambda$.
\end{proof}
Let $G' = \Contract(G, \atoms(\Tcal))$.  We now bound $w(E(G'))$.  We claim that every edge in $G'$ is either in $\Kcal$ or is incident to a vertex in $S$.  For if $\{u,v\} \in E(G')$ but $\{u,v\} \not \in \Kcal$, then for 
a cut $\Delta(Y)$ for $Y \in \Tcal$ with $\{u,v\} \in \Delta(Y)$ it must be the case that there is an $A \in \atoms(K)$ such that $u,v \in A$ and that for the $A_1, A_2 \in \atoms(K \cup Y)$ 
with $A = A_1 \cup A_2$, one of $A_1, A_2$ has size $< \frac{\beta d}{\tau}$.  This means that either $u$ or $v$ is small and so by \cref{clm:small}, $\{u,v\}$ is incident to $S$.  

The number of sets in $K$ is at most $\frac{\tau n}{\beta d}$ and for each $X \in K$ we have $w(\Delta(X)) \le \lambda + \varepsilon d \le (1+ \varepsilon) d$.  Thus 
we have that $w(\Kcal) \le (1+\varepsilon) \frac{\tau n}{\beta}$.  

Let us now bound the weight of edges incident to $S$.  As each vertex $v \in S$ has weight at least $\alpha w(v)$ amongst edges in $\Kcal$ we have that 
$\frac{\alpha}{2} \sum_{v \in S} w(v) \le w(\Kcal)$.  Thus overall we find 
\begin{align*}
w(E(G')) &\le w(\Kcal)\left( 1+ \frac{2}{\alpha}\right) \\
&\le (1+\varepsilon)(\alpha+2) \frac{\tau n}{\alpha \beta} \\
&\le \frac{68 \tau n}{(1-\varepsilon)^2} \enspace . \qedhere
\end{align*}
\end{proof}

The bound in \cref{thm:RSWweighted} is tight up to constant factors.
To see this, consider a cycle graph with uniform edge weight $\tau$.
Every edge participates in some minimum cut, and hence $G=G'$ and $w(E(G')) = \tau n$.


\section{Query-efficient quantum algorithm for minimum cut} \label{sec:query-efficient}
We first describe a query-efficient quantum algorithm to find the weight and shores of a minimum cut.
In \cref{sec:time-efficient} we make this algorithm time-efficient.
Our quantum query algorithm for minimum cut mainly relies on \cref{thm:RSWweighted}, and is inspired by a classical randomized algorithm for edge connectivity in the cut query model by Rubinstein, Schramm, and Weinberg (RSW) \cite{RSW18}.  
The RSW cut query algorithm is based on 4 subroutines whose input/output behavior we describe in Algorithms 1--4 below.  
For weighted graphs, we need an additional subroutine to compute the maximum weight of an edge in the graph which is stated in \cref{alg:maxweight}.
We describe all these subroutines in an abstract way to make it easy to (i) describe the time-efficient algorithm in the next section, and (ii) to instantiate this algorithm for other query models in the future.
We indicate oracle access to $G$ by square brackets and put the parameters explicitly given to the routines in parentheses.

\begin{algorithm}[H]
\caption{FindMinStar$[G](\delta)$}
\label{alg:mindeg}
 \hspace*{\algorithmicindent} \textbf{Input:} Oracle access to a weighted graph $G$, error parameter $\delta$. \\
 \hspace*{\algorithmicindent} \textbf{Output:} With probability at least $1-\delta$ output $v \in \argmin_{u \in V} w(\Delta_G(\{u\}))$ and 
 $d_{\min} = \min_{u \in V} w(\Delta_G(\{u\}))$.
\end{algorithm}

\begin{algorithm}[H]
\caption{Cut-Sparsifier$[G](\varepsilon, \delta)$}
\label{alg:sparsify}
 \hspace*{\algorithmicindent} \textbf{Input:} Oracle access to a weighted graph $G$, sparsifier accuracy parameter $\varepsilon$, error parameter $\delta$. \\
 \hspace*{\algorithmicindent} \textbf{Output:} With probability at least $1-\delta$ output an integer-weighted $\varepsilon$-cut sparsifier $H$ of $G$ with $\tO(n/\epsilon^2)$ edges.
\end{algorithm}

\begin{algorithm}[H]
\caption{LearnCutAtoms$(H,\lambda, \delta)$}
 \hspace*{\algorithmicindent} \textbf{Input:} Adjacency array description of $H$, cut threshold $\lambda$, and error parameter $\delta$. \\
 \hspace*{\algorithmicindent} \textbf{Output:}
 Define the set $\Tcal = \{ X: |X|, |\overline{X}| \ge 2, w(\Delta_H(X)) \le \lambda \}$.
 With probability at least $1-\delta$ output $\atoms(\Tcal)$.
\end{algorithm}

\begin{algorithm}[H]
\caption{LearnContraction$[G](\Pcal, M, \delta)$}
\label{alg:contract}
 \hspace*{\algorithmicindent} \textbf{Input:} Oracle access to a weighted graph $G$, a partition $\Pcal$ of $V(G)$, a natural number $M$, and error parameter $\delta$. \\
 \hspace*{\algorithmicindent} \textbf{Output:} Let $G' = \Contract(G,\Pcal)$.  With probability at least $1-\delta$ return $G'$ if the number of edges of $G'$ is at most $M$, 
 and otherwise return $\mathrm{NULL}$.
\end{algorithm}

\begin{algorithm}[H]
\caption{FindMaxWeight$[G](\delta)$}
\label{alg:maxweight}
 \hspace*{\algorithmicindent} \textbf{Input:} Oracle access to a weighted graph $G$, error parameter $\delta$. \\
 \hspace*{\algorithmicindent} \textbf{Output:} With probability at least $1-\delta$ output $\tau$, the maximum weight of an edge of $G$.
\end{algorithm}

We combine these subroutines in \cref{alg:mincut} to give a template for solving the minimum cut problem in an abstract query model.
\begin{algorithm}[H]
\caption{Query algorithm for minimum cut}
\label{alg:mincut}
 \hspace*{\algorithmicindent} \textbf{Input:} Oracle access to a weighted graph $G$ \\
 \hspace*{\algorithmicindent} \textbf{Output:} $\lambda(G)$ and the shores of a minimum cut of $G$.
\begin{algorithmic}[1]
\State $(v,d_{\min}) \gets$ FindMinStar$[G](\frac{1}{20})$.
\State $\tau \gets$ FindMaxWeight$[G](\frac{1}{20})$.
\State $H = (V,w') \gets$ Cut-Sparsifier$[G](\frac{1}{100}, \frac{1}{20})$.
\State Compute $\lambda(H)$.
\State $\Pcal = \{ S_1,\dots,S_k \} \gets$ LearnCutAtoms$(H,(1+\frac{1}{100})\lambda(H),\frac{1}{20})$.
\State $G' \gets$ LearnContraction$[G](\Pcal, 100 \tau n, \frac{1}{20})$.  If $G' = \mathrm{NULL}$ then abort.
\State Compute the weight $\lambda(G')$ and shores $(Y,V(G') \setminus Y)$ of a minimum cut in $G'$.
\State If $d_{\min} \le \lambda(G')$ output $(d_{\min},(\{v\},V\backslash\{v\}))$. Otherwise, let $Z = \cup_{S_i \in Y} S_i$ and output $(\lambda(G'), (Z, \overline{Z}))$.
\end{algorithmic}
\end{algorithm}

\begin{theorem}
\label{thm:rsw_alg}
Let $G$ be a weighted graph with $n$ vertices, minimum edge weight at least 1, and maximum edge weight $\tau$. 
\cref{alg:mincut} finds the weight and shores of a minimum cut of $G$ with probability at least $3/4$.  The number of queries of the algorithm is the sum of the number of queries 
of the subroutines FindMinStar$[G](\frac{1}{20})$, FindMaxWeight$[G](\frac{1}{20})$, Cut-Sparsifier$[G](\frac{1}{100}, \frac{1}{20})$, and LearnContraction$[G](\Pcal, 100 \tau n, \frac{1}{20})$.
\end{theorem}

\begin{proof}
Queries to the input graph $G$ are only made in steps $1,2,3,$ and $6$. This gives the statement about the complexity of the algorithm.  

Next let us deal with the error probability.  With probability at least $16/20$ steps 1--5 return correctly by the definition of these subroutines and 
the error parameter provided.  Let us now assume this is the case.   Then $H =(V,w')$ is a valid $\varepsilon$-sparsifier of $G$ for $\varepsilon = 1/100$.  
Let $X \in \Tcal$.  Then we have 
\[
w(\Delta_G(X)) \le (1+\varepsilon) w'(\Delta_H(X)) \le (1+\varepsilon)(1+3\varepsilon)\lambda(H) \le (1+\varepsilon)^2(1+3\varepsilon)\lambda(G) \enspace .
\]
We have $(1+\varepsilon)^2(1+3\varepsilon) \le \frac{11}{10}$ by the choice of $\varepsilon$, and so $w(\Delta_G(X)) \leq \frac{11}{10} \lambda(G) \leq \lambda(G) + \frac{1}{10} d_{\min}$ 
since $\lambda(G) \le d_{\min}$.  Thus by \cref{thm:RSWweighted}, the total weight of edges in $\Contract(G, \Pcal)$ will be at most $100\tau n$.
As we assume the minimum weight of an edge is at least $1$, the number of edges in $\Contract(G, \Pcal)$ will also be at most $100\tau n$. 
Hence except with probability at most $1/20$, LearnContraction will correctly return $\Contract(G, \Pcal)$ in step~5.

We have now argued that with probability at least $3/4$ all subroutines will correctly return.  We now argue correctness assuming that this is the case.  
In this case, $G'$ will be a valid contraction of $G$ and so $\lambda(G') \ge \lambda(G)$.  Thus if $\lambda(G)$ is achieved by a star cut the 
algorithm will return correctly.  

Let us now assume that $d_{\min} > \lambda(G)$ and let $\Delta_G(X)$ be a non-star cut with $w(\Delta_G(X)) = \lambda(G)$.  We have 
\[
w'(\Delta_H(X)) \le (1+\varepsilon) w(\Delta_G(X)) = (1+\varepsilon) \lambda(G) \le \frac{1+\varepsilon}{1-\varepsilon} \lambda(H) \le (1+3\varepsilon) \lambda(H)\enspace ,
\]
where the last step holds as $\varepsilon \le \frac{1}{3}$.  This means $X \in \Tcal$ and therefore no edge of $\Delta_G(X)$ will be 
contracted in $G' = \Contract(G,\Pcal)$.  Thus $\lambda(G') \le \lambda(G)$ and as the edge connectivity cannot decrease in a contraction in fact $\lambda(G') = \lambda(G)$.
Hence the algorithm returns correctly in step~8.
\end{proof}

\begin{lemma} \label{lem:compl-subroutines}
Let $G = (V,w)$ be a weighted graph with $n$ vertices and $m$ edges.
Subroutines FindMinStar$[G](\frac{1}{20})$, FindMaxWeight$[G](\frac{1}{20})$, Cut-Sparsifier$[G](\frac{1}{100}, \frac{1}{20})$ can be implemented by a quantum algorithm 
with query and time complexity  $\tO(n^{3/2})$ in the adjacency matrix and $\tO(\sqrt{mn})$ in the adjacency array model.  

LearnContraction$[G](\Pcal, 100 \tau n, \frac{1}{20})$ can be implemented by a quantum algorithm with query and time complexity  
$\tO(n^{3/2}\sqrt{\tau})$ in the adjacency matrix and $\tO(\sqrt{mn\tau})$ in the adjacency array model.
\end{lemma}

\begin{proof}
First note that in the adjacency array model we may assume that $m \ge n$. Otherwise, $\sqrt{mn} \ge m$ and we can perform each task classically in $\tO(m)$ time and queries.
We consider each of the subroutines in turn:

\textbf{FindMinStar$[G](\frac{1}{20})$:} In the adjacency matrix model we can compute $w(\Delta_G(\{v\})$ with $n-1$ classical queries to the adjacency matrix.  
We can compose this with quantum minimum finding to find the minimum weight of a star cut and a vertex realizing this in query and time complexity $\tO(n^{3/2})$ by \cref{thm:qmin}.

In the adjacency array model we first classically query the degrees of all the vertices with $n$ queries.  In a simple graph this suffices to determine the minimum weight of a star cut.
In a weighted graph we continue as follows.  For $1 \leq \ell \leq \ceil{\log n}$, define the bucket $B_\ell \subseteq V$ as the subset of nodes $v$ that have degree in $[2^{\ell-1}, 2^\ell)$.
As the sum of the degrees is $2m$ we have that $|B_\ell| \leq 2m/2^{\ell-1}$.
Finding the minimum $\min_{v \in B_\ell} w(\Delta_G(\{v\}))$ over a single bucket has quantum query and time complexity $\tO(\sqrt{mn})$: we can compute $w(\Delta_G(\{v\}))$ for a single $v \in B_\ell$ using at most $2^\ell$ classical queries, and 
then do quantum minimum finding over the $|B_\ell| \leq 2m/2^{\ell-1}$ nodes in $B_\ell$.
This has total query and time complexity $\tO(2^\ell \sqrt{2m/2^{\ell-1}}) \in \tO(\sqrt{m 2^\ell}) \in \tO(\sqrt{mn})$.
We do this for each of the $\ceil{\log n}$ buckets and we output the minimum overall weight and a vertex realizing this.
This yields a total time and query complexity $\tO(\sqrt{mn})$.

\textbf{FindMaxWeight$[G](\frac{1}{20})$:}
This amounts to finding the maximum of a set of $n^2$ numbers in the adjacency matrix model, or $m$ numbers in the adjacency list model.
By \cref{thm:qmin} this has query and time complexity $\tO(n)$ and $\tO(\sqrt{m})$, respectively.

\textbf{Cut-Sparsifier$[G](\frac{1}{100}, \frac{1}{20})$:} A $\frac{1}{100}$-cut sparsifier with $\tO(n/\epsilon^2)$ edges can be constructed with high probability in query and time complexity $\tilde O(n^{3/2})$ in the adjacency matrix model or $\tilde O(\sqrt{mn})$ in the adjacency array model by \cref{thm:qsparsifier}.  

\textbf{LearnContraction$[G](\Pcal, 100\tau n, \frac{1}{20})$:} 
First we handle a trivial case.  If $\tau \ge n$ then we can classically learn the input in time $n^2 = O(n^{3/2}\sqrt{\tau})$ in the adjacency matrix model and 
time $m = O(\sqrt{mn\tau})$ in the adjacency array model.  Thus we can assume $\tau < n$.  

First we do the adjacency matrix case.  Let $x \in \R^{\binom{n}{2}}$ be a vector whose entries are labeled by elements of $V^{(2)}$ and where $x(e) = w(e)$ if the endpoints of $e$ are in distinct elements of $\Pcal$ and $x(e) = 0$ otherwise.  
A query to an entry of $x$ can be answered with one query to the adjacency matrix of $G$.  Let $\hat{x} \in \{0,1\}^{\binom{n}{2}}$ be defined by $\hat{x}(e) = 1$ if $w(e) > 0$ and $\hat{x}(e) = 0$ otherwise.  We can also answer a query to $\hat{x}$ 
with one query to the adjacency matrix of $G$.  By \cref{thm:qsearch} in query and time complexity 
$\tO(n^{3/2}\sqrt{\tau})$ in the adjacency matrix model we can with probability at least $9/10$ 
output $\hat{x}$ if $|\hat{x}| \le 100 \tau n$ and otherwise output NULL.  We can then classically query $x$ in the non-zero locations of $\hat{x}$ with $100\tau n =O(n^{3/2} \sqrt{\tau})$ more classical queries to output $x$.  
This fulfils the specification of LearnContraction.

Similarly, in the adjacency array model let $x \in \R^{m}$ be labeled by entries of the adjacency array of $G$ and define $x(e) = w(e)$ if the endpoints of $e$ are in distinct elements of $\Pcal$ and 
$x(e) = 0$ otherwise.  Let $\hat{x}(e)=1$ if $x(e) >0$ and $\hat{x}(e)=0$ otherwise as before.
A query to an entry of $x$ or $\hat{x}$ can be answered with one query to the adjacency array of $G$.  Again by \cref{thm:qsearch}, in query and time complexity $\tO(\sqrt{mn \tau})$ in the adjacency array model we can with probability at least $9/10$ 
output $\hat{x}$ if $|\hat{x}| \le 100 \tau n$ and otherwise output NULL.  With $100 \tau n = O(\sqrt{mn\tau})$ more queries we can then output $x$.
\end{proof}

\begin{theorem}
\label{thm:main}
Let $G =(V,w)$ be an $n$-vertex weighted graph with $m$ edges and edge-weight ratio $\tau$.  There is a quantum algorithm that finds the 
weight and shores of a minimum cut of $G$ with probability at least $3/4$ after 
$\tO(n^{3/2}\sqrt{\tau})$ queries to the adjacency matrix of $G$ or $\tO(\sqrt{mn\tau})$ queries to the adjacency array.
\end{theorem}
\begin{proof}
First we use the minimization analogue of FindMaxWeight to find the minimum edge weight $\alpha$.  Then by normalizing by $1/\alpha$ we may assume that all 
edge weights are at least $1$ and apply \cref{thm:rsw_alg}.  The bound on the quantum query complexities then follows from \cref{lem:compl-subroutines}.
\end{proof}


\section{Time-efficient quantum algorithm for minimum cut} \label{sec:time-efficient}
In this section we describe a quantum algorithm for computing the weight of a minimum cut of a weighted graph with time complexity $\tO(\sqrt{mn\tau} + n^{3/2})$ in the adjacency array model and $\tO(n^{3/2}\sqrt{\tau})$ in the adjacency matrix model.
In the adjacency matrix model this is optimal up to polylogarithmic factors.
Our algorithm is a time-efficient implementation of \cref{alg:mincut}.  The running time of this algorithm is the sum of the running time of its 4 subroutines, and we have already 
analyzed the complexity of 3 of those subroutines in \cref{lem:compl-subroutines}.  Thus it now suffices to give a time-efficient implementation of the subroutine 
LearnCutAtoms, as formalized in the next lemma.

\begin{lemma}
\label{lem:kappa}
Let $\kappa(n)$ denote the maximum time complexity of a quantum algorithm for the subroutine LearnCutAtoms$(H,(1+\frac{1}{100})\lambda(H),\frac{1}{20})$ over weighted $n$-vertex graphs $H$ with $\tO(n)$ edges.
Let $G$ be a weighted graph with $n$ vertices, $m$ edges, and edge-weight ratio $\tau$.
There is a quantum algorithm to compute the weight and shores of a minimum cut of $G$ with probability at least $2/3$
that runs in time $\kappa(n) + \tO(\sqrt{mn\tau})$ in the adjacency array model and 
$\kappa(n) + \tO(n^{3/2}\sqrt{\tau})$ in the adjacency matrix model.
\end{lemma}
\begin{proof}
First we use minimum finding \cref{thm:qmin} to determine the minimum $\alpha$ and maximum $\beta$ edge weights with error probability at most $1/12$.  This requires time 
$\tO(\sqrt{m})$ in the adjacency array model and $\tO(n)$ in the adjacency matrix model and so will be low order to the time bounds stated in the lemma.  From $\alpha, \beta$ we compute the 
edge-weight ratio $\tau = \beta/\alpha$.  By multiplying all edge weights by $1/\alpha$ we may assume that the minimum edge weight is $1$ and 
the maximum edge weight is $\tau$.

If $\tau > m/n$ (in the adjacency array model) or $\tau > n$ (in the adjacency matrix model), then we simply run a randomized near-linear time algorithm (e.g., \cite{Karger00}) for calculating the weight and shores of a minimum cut of $G$.
This then takes time $\tO(m) \in \tO(\sqrt{mn\tau})$ in the array model and $\tO(n^2) \in \tO(n^{3/2}\sqrt{\tau})$ in the matrix model.
We can hence assume that $\tau \leq m/n$ in the array model and $\tau \leq n$ in the matrix model.

We use a quantum implementation of \cref{alg:mincut}.  By \cref{thm:rsw_alg} this algorithm has error probability at most $1/4$, thus our overall error probability will be at most $1/3$ as desired.  
For the running time it suffices to analyze the quantum time complexity of all 8 steps.
In \cref{lem:compl-subroutines} we show that the time complexity of steps 1--3 and 6 is $\tO(\sqrt{mn\tau})$ in the adjacency array model and $\tO(n^{3/2}\sqrt{\tau})$ in the adjacency matrix model.
For step 4, we can use a randomized near-linear time algorithm (e.g., \cite{Karger00}) for calculating the weight and shores of a minimum cut of $H$.  As $H$ has $\tO(n)$ edges this takes time $\tO(n)$.
In step 7, we compute the weight and shores of a minimum cut in $G'$ which has at most $100\tau n$ edges by the definition of LearnContraction.  This takes time $\tO(\tau n)$, which is $\tO(\sqrt{mn\tau})$ in the array model (by the assumption $\tau \leq m/n$) or $\tO(n^{3/2} \sqrt{\tau})$ in the matrix model (by the assumption $\tau \leq n$).
Finally, step 8 is trivial and the quantum time complexity of step 5 is exactly $\kappa(n)$.
\end{proof}

This section is hence devoted to proving the following theorem. 
\begin{restatable}{theorem}{cutatomsthm} \label{thm:learncutatoms}
Let $H$ be an $n$-vertex weighted graph with $m$ edges.
There is a quantum algorithm that implements LearnCutAtoms$(H,(1+\frac{1}{100})\lambda(H),\frac{1}{20})$ in time $\tO(m + n^{3/2})$.
\end{restatable}

In particular, \cref{thm:learncutatoms} implies that $\kappa(n) \in \tO(n^{3/2})$, and hence we find a time-efficient quantum algorithm.
\thmtime*

\begin{proof}
Follows from \cref{lem:kappa} and \cref{thm:learncutatoms}.  
\end{proof}

\subsection{Tools}

Our time efficient algorithm builds on a number of tools, which we first introduce here.

\subsubsection{2-respecting cuts and Karger's theorem}

In his seminal work on a near-linear time randomized algorithm for minimum cut \cite{Karger00}, Karger combined sparsification with the notion of \emph{tree-respecting cuts}.
Consider an $n$-vertex graph $G = (V,w)$, a spanning tree $T$ and a cut with shore $X$.
We say that the cut \emph{2-respects} $T$ if it cuts at most 2 edges of $T$, i.e., $|\Delta_T(X)| \leq 2$, and \emph{strictly 2-respects} $T$ if $|\Delta_T(X)| = 2$.  Note that 
the set of cuts which 2-respect $T$ depends only on $E(T)$ and not the weight of edges in $T$.  Note also that there are $n-1 + \binom{n-1}{2} = \binom{n}{2}$ cuts that 2-respect $T$.

Karger proved that we can efficiently construct a set of $O(\log n)$ spanning trees of $G$ such that every minimum cut of $G$ will 2-respect a constant fraction of them.
This effectively reduces the exponentially large search space for finding a minimum cut to the set of merely $O(n^2 \log n)$ cuts that 2-respect one of the spanning trees.
For our purpose, we will use these spanning trees as an efficient representation of the near-minimum cuts of the graph.  For this, we need a slight generalization of Karger's theorem 
on tree-respecting cuts.  This shows we can efficiently find $O(\log n)$ spanning trees such that any $(1+1/16)$-near-minimum cut $2$-respects a constant fraction of them, while Karger's statement was 
only for minimum cuts.  This only requires a minor modification of Karger's proof, but for completeness we provide a proof in \cref{app:karger}. 

Throughout this section we will use the phrase ``with high probability'' to mean with probability at least $1-1/n^c$ for an arbitrary constant $c$.
\begin{restatable}[{\cite[Theorem 4.1]{Karger00}}]{theorem}{kargerthm} \label{thm:karger}
Let $G = (V,w)$ be a weighted graph with $n$ vertices and $m$ edges.  There is a randomized algorithm that in time 
$O(m \log^2(n) + n\log^4(n))$ time constructs a set of $O(\log n)$ spanning trees such that every $(1+1/16)$-near minimum cut of $G$ 2-respects $1/4$ of them with high probability.
\end{restatable}
Karger states the runtime of the algorithm in this theorem as $O(m + n \log^3(n))$, but we opt for a simpler proof rather optimizing log factors.

\subsubsection{Data structures}
We will frequently need to refer to a 2-respecting cut both by its shores and the edges of the tree it cuts.  We develop some notation to make this easier.
\begin{definition}[Notation for 2-respecting cuts]
Let $T$ be a tree on vertex set $V$ with root $r$.  Define $N(T) = E(T) \cup E(T)^{(2)}$.
For $f \in N(T)$ define $\shore(f)$ to be the set $X \subseteq V$  such that $\Delta_T(X) = f$ and $X$ does not contain $r$.
For $X \subseteq V$ such that $|\Delta_T(X)| \le 2$, 
let $\edges(X) = \Delta_T(X)$.  We overload both these notations to sets so that $\shore(Q) = \{\shore(f): f \in Q\}$ for $Q \subseteq N(T)$ and similarly 
$\edges(\Tcal) = \{\Delta_T(X): X \in \Tcal\}$ for a set $\Tcal$ of shores of 2-respecting cuts of $T$.  
\end{definition}

With some preprocessing time, we can efficiently evaluate the weight of 2-respecting cuts.  The following lemma is very useful.
\begin{lemma}[{\cite[Lemma 1]{GMW20c}}] \label{lem:eval-cuts}
Given a weighted graph $G = (V,w)$ with $n$ vertices and $m$ edges, and a spanning tree $T$ of $G$, we can construct in $O(m \log n)$ time a data structure that, for any $f \in N(T)$,
reports the weight $w(\Delta_G(\shore(f)))$ of the corresponding 2-respecting cut in $O(\log n)$ time.
\end{lemma}

Another data structure that we use is based on the \emph{Euler tour technique} \cite{TV84,HK95}.
This is a way of representing a tree that is useful to access and modify data in subtrees.
Consider an undirected tree $T=(V_T,E_T)$ with root $r \in V_T$.
To $T$ we associate the directed graph $\vec T = (V_T, \vec E_T)$ obtained by replacing every edge in $E_T$ by a pair of directed edges in opposite directions.
Now let $\Ecal_T \in (\vec E_T)^{2(n-1)}$ denote an Euler tour in $\vec T$, starting and ending in root $r$.
$\Ecal_T$ is a sequence of $2(n-1)$ edges as each directed edge is traversed exactly once.

For every node $u$ in $V_T$, let $f(u)$ be the index in $\Ecal_T$ of the edge that points toward $u$, and let $\ell(u)$ be the index of the last edge that points toward $u$.
Now if $T(u)$ is the subtree of $T$ induced by vertex $u$ and all of its descendants, then the subsequence of $\Ecal_T$ starting at $f(u)$ and ending at $\ell(u)$ (both included) is an Euler tour representation of $T(u)$.
Hence any subtree corresponds to a subsequence of $\Ecal_T$.
We can use this to prove the lemma below, which will be useful to compute $\atoms(\Tcal)$ from a set $\Tcal$ of shores of cuts that 2-respect a given tree.

Given a tree whose nodes have some key value, we call a \emph{subtree-add} the increasing or decreasing of the key value in a subtree by some fixed amount.

\begin{lemma} \label{lem:euler-tree}
Let $T=(V_T,E_T)$ be a tree with key values $\{k_u \mid u \in V_T\}$ of $O(\log n)$ bits.
There is a data structure that implements $M$ subtree-adds in time $\tO(n + M)$.
\end{lemma}
\begin{proof}
Fix a root node $r$.
Represent $T$ by an Euler tour $\Ecal_T \in (\vec E_T)^{2(n-1)}$ and define $f(u),\ell(u)$ for each $u \in V$ as above.
Associate to $\Ecal_T$ a list $A$ of length $2(n-1)$ to store the key values, setting $A(i) = k_u$ if the $i$-th entry of $\Ecal_T$ is an edge whose tail is $u$.
Adding value $\alpha$ to the keys of nodes in subtree $T(u)$ amounts to adding $\alpha$ to every entry in the subsequence in $A$ starting with $f(u)$ and ending with $\ell(u)$ (both included).
Call such an operation $\texttt{ADD}(\alpha,f(u),\ell(u))$.

To implement $M$ \texttt{ADD} operations, create a second emtpy list $B$ with length $2(n-1)$.
For every operation $\texttt{ADD}(\alpha,f(u),\ell(u))$, set $B(f(u)) = B(f(u)) + \alpha$ and if $\ell(u) < 2(n-1)$ set $B(\ell(u)+1) = B(\ell(u)+1) - \alpha$.
Now do a partial sum transformation of $B$:
\begin{algorithmic}[1]
\State
Create list $s_B$ of length $2(n-1)$ with $s_B(1) = B(1)$ and $s_B(i) = 0$ for all $i \in [2,2(n-1)]$.
\For{$i = 2,3,\dots,2(n-1)$}
\State
Set $s_B(i) = s_B(i-1) + B(i)$.
\EndFor
\end{algorithmic}
In total this has time complexity $\tO(n + M)$ (assuming $\tO(1)$ cost for arithmetic operations).
The final key values are now given by setting $k_u = A(f(u)) + B(f(u))$.
\end{proof}

\subsection{Generating set for a single tree} \label{sec:single-tree}
Let $G = (V,w)$ be an $n$-vertex weighted graph and $T$ be a spanning tree of $G$.  Let $Q \subseteq E(T)^{(2)}$ and 
$\Mcal = \shore(Q)$.  In words, $\Mcal$ is an arbitrary set of shores of cuts that \emph{strictly} 2-respect $T$.  
The next lemma gives an explicit generating set $\Scal$ for $\atoms(\Mcal)$ with $|\Scal| \le n-2$.  We first make a definition
that will be used throughout this section.

\begin{definition}[separate]
Let $V$ be a finite set and $X \subseteq V$.  For $u,v \in V$ we say that $X$ separates $u,v$ if exactly one of 
them is in $X$.
\end{definition}

\begin{lemma}
\label{lem:spanning}
Let $T$ be a tree on a vertex set $V$ of cardinality $n$.  Let $\Mcal \subseteq 2^V$ be a set of shores that strictly 2-respect $T$ and let 
$Q = \edges(\Mcal)$.  Define the graph $L = (E(T), Q)$ and let $F$ be a spanning forest of $L$.  
Then $\Scal = \shore(E(F))$ is a generating set for $\atoms(\Mcal)$.
\end{lemma}

\begin{proof}
Clearly $E(F) \subseteq Q$ thus $\Scal \subseteq \Mcal$.  This means that $\atoms(\Mcal)$ is a refinement of $\atoms(\Scal)$.  
Thus to show $\atoms(\Scal) = \atoms(\Mcal)$ it suffices to show that any $u,v \in V$ that are in different sets of $\atoms(\Mcal)$ are also in 
different sets of $\atoms(\Scal)$.  
 
The key fact we need is that if $\Delta_T(X) = \{e,e'\}$ then $X$ separates $u,v$ iff exactly one of $e,e'$ is on the 
path from $u$ to $v$ in $T$.  Suppose that $u,v$ are in different sets of $\atoms(\Mcal)$, that is there is an $X \in \Mcal$ 
which separates them.  Say that $\Delta_T(X) = \{e_\rmin,e_\rmout\}$ where $e_\rmin$ is on the $u-v$ path in 
$T$ and $e_\rmout$ is not.  Then $\{e_\rmin,e_\rmout\} \in Q$ and therefore there must be a path between $e_\rmin$ and $e_\rmout$ in the spanning forest 
$F$.  Let $(e_0,e_1, e_2, \ldots, e_k)$, where $e_0 = e_\rmin, e_k = e_\rmout$, be the sequence of vertices on this path in $F$.  As $e_\rmin$ is on the $u-v$ path in $T$ and $e_\rmout$ is 
not, there must be consecutive vertices $e_i,e_{i+1}$ where $e_i$ is on the $u-v$ path in $T$ and $e_{i+1}$ is not.  As $\{e_i, e_{i+1}\} \in E(F)$ there is 
an $X \in \Scal$ which separates $u$ and $v$.  
\end{proof}

\begin{lemma}
\label{lem:q_genset}
Let $G = (V,w)$ be an $n$-vertex weighted graph with $m$ edges and $T$ a spanning tree of $G$.  For a real number 
$\alpha \ge 1$, let $\Tcal = \{X \subseteq V: w(\Delta_G(X)) \le \alpha \lambda(G), |\Delta_T(X)| \le 2\}$.  
There is a quantum algorithm that outputs with high probability a set $Q \subseteq N(T)$ in time $\tO(m + n^{3/2})$ such that $|Q| \le 2n-3$ and $\Scal = \shore(Q)$
is a generating set for $\atoms(\Tcal)$.  
\end{lemma}

\begin{proof}
Let $\Tcal_1 = \{X \in \Tcal: |\Delta_T(X)| =1\}$ and $\Tcal_2 = \{X \in \Tcal: |\Delta_T(X)| =2\}$. Let $Q_1 = \edges(\Tcal_1)$ and $Q_2 = \edges(\Tcal_2)$.  
Let $F$ be a spanning tree for $L = (E(T), Q_2)$.  By \cref{lem:spanning}, $\Rcal = \shore(E(F))$ is a generating set for 
$\atoms(\Tcal_2)$ and $|\Rcal| \le n-2$ as $F$ is a spanning tree of an $n-1$-vertex graph.  Thus by \cref{prop:atom_cup}, $\Scal = \Tcal_1 \cup \Rcal$ is a generating set for $\Tcal$ of size at most $2n-3$.  
Thus taking $Q = Q_1 \cup E(F)$ satisfies the conditions of the lemma.

Now we must show how to efficiently output $Q$.  We can first run a near-linear time classical randomized algorithm to compute $\lambda(G)$ \cite{Karger00}.  We then in near-linear time set up the 
data structure given by \cref{lem:eval-cuts}.  For an $f \in N(T)$ this lets us check in time $O(\log(n))$ if $f \in Q_1 \cup Q_2$.  We can then cycle over the edges $e\in E(T)$ to create the set $Q_1$ classically in time 
$\tO(n)$.  It now remains to construct a spanning tree of $L = (E(T), Q_2)$.
For any $f \in E(T)^{(2)}$ we can use the data structure to check in $O(\log n)$ time if $f \in Q_2$.
This gives us adjacency matrix access to $L$ with $O(\log n)$ overhead for each query.  
Now we can use the quantum algorithm from \cite{DHHM06} that with high probability outputs a spanning forest of an $n$ vertex graph in the adjacency matrix model with $\tO(n^{3/2})$  queries and time.  
Thus we can use this algorithm to construct a spanning forest $F$ of $L$.  We then output $Q = Q_1 \cup E(F)$ as desired.
\end{proof}

Now we have an implicit representation $\edges(\Scal)$ of a generating set $\Scal$ for $\atoms(\Tcal)$, where $\Tcal$ is the set of near-minimum cuts of a graph $G$ that 2-respect a tree $T$.  
What we need, however, is to actually output $\atoms(\Tcal)$.
In the following lemma we show how to do this efficiently by combining random hashing with Euler tour trees.

\begin{lemma} \label{lem:2-resp-part}
Let $T$ be a tree on a vertex set $V$ of size $n$, $Q \subseteq N(T)$, and $\Scal = \shore(Q)$.  
Given input $Q$ there is a classical algorithm that with probability at least $1-1/n$ outputs $\atoms(\Scal)$ in time $\tO(n + |Q|)$.
\end{lemma}
\begin{proof}
Let $M$ be a large integer to be chosen later and consider the following algorithm.
Pick $\ell \in \Z_M$ uniformly at random and give every vertex $u \in V$ the key value $k_u = \ell$.
For every $f \in Q$, do:
\begin{itemize}
\item
Pick $\ell \in \Z_M$ uniformly at random and set $k_u = k_u + \ell \;(\bmod M)$ for all $u \in \shore(f)$.
\end{itemize}
Now if $u$ and $v$ are in the same set of $\atoms(\Scal)$, that is no set of $\Scal$ separates them, then $k_u = k_v$.
On the other hand, if $u$ and $v$ are in different sets of $\atoms(\Scal)$ then there is some $f \in Q$ such that $u \in \shore(f)$ and $v \not \in \shore(f)$, or vice versa. In this case,
$k_u$ and $k_v$ are pairwise independent 
and distributed uniformly at random in $\Z_M$.
Hence $k_u = k_v$ with probability $1/M$.
Taking a union bound over all pairs $u,v$, we see that with probability at least $1-\binom{n}{2}/M$ we have that $k_u \neq k_v$ for all $u,v$ in different sets of $\atoms(\Scal)$.
If we set $M = n^3$ and we let $\Pcal(\{k_u\})$ denote the partition induced by gathering nodes with the same key value, then $\Pcal(\{k_u\}) = \atoms(\Scal)$ with probability at least $1 - 1/n$.

The cost of actually implementing this algorithm is dominated by sequentially updating for every $f \in Q$ the key value for all nodes in $\shore(f)$.
This amounts to changing the key value in at most 2 subtrees of $T$:
\begin{itemize}
\item
If $f = e \in E(T)$, then $\shore(f)$ is the subtree $T(u)$ of some node $u$ and we have to change the key value in $T(u)$.
\item
If $f = \{e,e'\} \in E(T)^{(2)}$, then we distinguish two cases.
If one of the two cut edges is a descendant of the other then $\shore(f)$ is of the form $T(u) \setminus T(v)$ for two nodes $u,v \in V$.
In this case we can update the key values by adding $\ell$ to $T(u)$ and subtracting $\ell$ from $T(v)$.
If neither of the edge is a descendant of the other then $\shore(f)$ is of the form $T(u) \cup T(v)$, and we can update the key values by adding $\ell$ to $T(u)$ and $T(v)$.
\end{itemize}
In \cref{lem:euler-tree} we show how to change the key values in $|Q|$ subtrees in total time $\tO(n + |Q|)$ using Euler tour trees.
\end{proof}

We can now put all these pieces together into the following algorithm.
\begin{algorithm}[H]
\caption{Algorithm for finding atoms of the shores of 2-respecting near-minimum cuts}
\label{alg:partition-tree}
 \hspace*{\algorithmicindent} \textbf{Input:} Explicit description of $G=(V,w)$, a spanning tree $T$ of $G$, a real number $\alpha \ge 1$. \\
 \hspace*{\algorithmicindent} \textbf{Output:} $\atoms(\Tcal)$ where $\Tcal = \{X: w(\Delta_G(X)) \le \alpha \lambda(G) \mbox{ and } |\Delta_T(X)| \le 2\}$.
\begin{algorithmic}[1]
\State
Compute $\lambda(G)$.
\State
Create data structure as in \cref{lem:eval-cuts} for evaluating the weight of cuts in $G$ that 2-respect $T$.
\State
Compute $Q$ such that $\shore(Q)$ is a generating set for $\atoms(\Tcal)$ by \cref{lem:q_genset}.
\State
Use \cref{lem:2-resp-part} to find and return $\atoms(\shore(Q)) = \atoms(\Tcal)$.
\end{algorithmic}
\end{algorithm}

\begin{lemma} \label{lemma:2-respect}
Let $G=(V,w)$ be an $n$-vertex weighted graph with $m$ edges and $T$ a spanning tree of $G$.  Let $\alpha \ge 1$ be a real number and
$\Tcal = \{X: w(\Delta_G(X)) \le \alpha \lambda(G) \mbox{ and } |\Delta_T(X)| \le 2\}$.  
\cref{alg:partition-tree} outputs $\atoms(\Tcal)$ with high probability and can be implemented by a quantum algorithm in time $\tO(m + n^{3/2})$.
\end{lemma}

\subsection{Time-Efficient quantum algorithm for LearnCutAtoms}
We now describe a time-efficient quantum algorithm for outputting $\atoms(\Tcal)$, where $\Tcal$ is the set of shores of all $(1+1/100)$-near-minimum cuts of a weighted graph $H$.
This algorithm combines Karger's tree packing \cref{thm:karger} with the algorithm that produces the atoms of shores of cuts that 2-respect a tree from the previous section (\cref{lemma:2-respect}).

\begin{algorithm}[H]
\caption{LearnCutAtoms$(H,\lambda,\delta)$}
\label{alg:partition}
 \hspace*{\algorithmicindent} \textbf{Input:} Explicit description of an $n$-vertex weighted graph $H = (V,w)$ with $m$ edges, a cut threshold $\lambda \le (1+1/16) \lambda(H)$, and an error parameter $\delta$. \\
 \hspace*{\algorithmicindent} \textbf{Output:} $\atoms(\Tcal)$ where $\Tcal = \{X \subseteq V: w(\Delta_G(X)) \le \lambda\}$.
\begin{algorithmic}[1]
\State
Construct set of $K \in O(\log n)$ spanning trees $\{T_i\}$ using \cref{thm:karger}.
\For{$i = 1,2,\dots,K$}
\State
Use \cref{alg:partition-tree} to find $\atoms(\Tcal_i)$ where $\Tcal_i = \{X \subseteq V: w(\Delta_G(X)) \le \lambda \mbox{ and } |\Delta_{T_i}(X)| \le 2\}$.
\EndFor
\State
Output $\atoms(\cup_i \atoms(\Tcal_i))$.
\end{algorithmic}
\end{algorithm}

\cutatomsthm*

\begin{proof}
We use \cref{alg:partition}.  First let us argue correctness.  As $\lambda \le (1+16)\lambda(H)$, by \cref{thm:karger} for every $X \in \Tcal$ 
there will be a tree $T_i$ such that $\Delta_{T_i}(X) \le 2$.  This means that $\Tcal = \cup_{i=1}^K \Tcal_i$.  Hence 
$\atoms(\Tcal) = \atoms(\cup \Tcal_i) = \atoms(\cup \atoms(\Tcal_i))$.  By \cref{lemma:2-respect}, step~(3) correctly outputs $\atoms(\Tcal_i)$ 
for $i = 1, \ldots, K$ with high probability, and thus step~(5) will output $\atoms(\Tcal)$ with high probability.

Now let us analyze the complexity.  Step~(1) can be done in $\tO(m)$ time by a classical randomized algorithm by \cref{thm:karger}.  Step~(3) can 
be done by a quantum algorithm in time $\tO(m + n^{3/2})$ by \cref{lemma:2-respect}, and thus the for loop has the same time bound as $K = O(\log n)$.

Finally, we need to explain how to (classically) implement step~(5).  First we give every node $v \in V$ a key value $k_v = 0$.
Then, for each $i=1,\ldots, K$, we iterate over the node set and append a $\log n$-bit string to the key value of every node, indicating the component of $\atoms(\Tcal_i)$ of which it is part.
At the end of this routine every node has a $O(\log^2 n)$-bit key value that indicates its component in $\atoms(\Tcal)$.  The total runtime for this step is $\tO(n)$.  Thus overall 
the running time is $\tO(m + n^{3/2})$.
\end{proof}


\section{Lower bounds} \label{sec:lower-bounds}
In this section we present lower bounds on the complexity of edge connectivity and weighted minimum cut.

First we describe some existing lower bounds for the case of simple graphs.
Let $\CON_n$ be the problem of deciding if an input \emph{simple} graph on $n$ vertices is connected or not.  This is a special case of edge connectivity, where one wants 
to decide if the edge connectivity is zero or positive.  D\"{u}rr, Heiligman, H{\o}yer and Mhalla \cite{DHHM06} proved the following quantum query lower bounds on the complexity of $\CON_n$.
\begin{theorem}[\cite{DHHM06}]
\label{thm:durr_lower}
The bounded-error quantum query complexity of $\CON_n$ is $\Theta(n^{3/2})$ in the adjacency matrix model and 
$\Theta(n)$ in the adjacency array model.  
\end{theorem}
\noindent
This theorem shows that, in the adjacency matrix model, \cref{thm:main} is tight up to polylogarithmic factors for simple graphs.  For the adjacency array model 
there is still a gap between the $\Omega(n)$ lower bound from \cref{thm:durr_lower} and the $\tilde O(\sqrt{mn})$ upper bound for simple graphs given by \cref{thm:main}.

For the minimum cut problem in a weighted graph we prove separate and distinct lower bounds for the adjacency matrix model and the adjacency array model.
All our lower bounds essentially follow by forcing the algorithm to solve a counting problem in order to compute the weight of a minimum cut.
We then use the following theorem by Nayak and Wu that gives a lower bound on the quantum query complexity of exact counting.

\begin{theorem}[{\cite[Corollary 1.2]{NW99}}]
\label{thm:thresh}
Let $k,N \in \N$ with $2k+1 \le N$.  Assume query access to $x \in \{0,1\}^N$ with the promise that $|x| = k+1$ or $|x| = k-1$.  Any quantum algorithm that correctly decides whether $|x| = k+1$ or $|x| = k-1$ with probability at least $2/3$ must make $\Omega(\sqrt{N k})$ queries.
\end{theorem}

\subsection{Adjacency matrix model}

In the adjacency matrix model we show that for any integer $1 \le \tau \le (\floor{n/2}-1)/2$, in the worst case $\Omega( n^{3/2} \sqrt{\tau})$ adjacency matrix queries are needed to compute the weight of a minimum cut of a graph with edge weights in 
$\{1,\tau\}$.
This matches the upper bound in \cref{thm:main}, and hence settles the quantum query complexity of weighted minimum cut in the adjacency matrix model.
For $\tau = 1$ this reproduces the aforementioned $\Omega(n^{3/2})$ bound which follows from \cite{DHHM06}.

\begin{theorem}
\label{thm:lower}
Let $n,\tau \in \N$ satisfy $1 \le \tau \le (\floor{n/2}-1)/2$.  There is a family of $n$-vertex graphs $\Gcal$ all of which have edge weights in $\{0,1,\tau\}$ 
such that any quantum algorithm that for every graph $G \in \Gcal$ computes with probability at least $2/3$ the weight of a minimum cut in $G$ must make 
$\Omega(n^{3/2}\sqrt{\tau})$ queries in the adjacency matrix model.  Similarly, any quantum algorithm that for every graph $G \in \Gcal$ computes with 
probability at least $2/3$ the shores $(X,\overline{X})$ of a cut realizing the minimum weight must make $\Omega(n^{3/2}\sqrt{\tau})$ queries in the adjacency matrix model.
\end{theorem}

\begin{proof}
Let $V$ be an $n$-element set and partition $V$ into disjoint sets $V = V_0 \sqcup V_1$ where $|V_0| = \floor{n/2}$, $|V_1| = \ceil{n/2}$.  Choose a distinguished vertex $v_0 \in V_0$, and let $V_0' = V_0 \setminus \{v_0\}$. 
Let $N = |V_0' \times V_1|$ and let $g : V_0' \times V_1 \rightarrow [N]$ be a bijection.  For every $x \in \{0,1\}^N$ we define a weighted graph $G_x = (V,w_x)$ where
\begin{itemize}
\item
$w_x(\{u,v\}) = \tau$ if $u,v \in V_0$ or $u,v \in V_1$,
\item
$w_x(\{u,v\}) = x(g(\{u,v\}))$ if $(u \in V_0', v \in V_1)$ or $(u \in V_1, v \in V_0')$,
\item
$w_x(\{u,v\}) = 0$ otherwise.
\end{itemize}
Let $k = \tau (\floor{n/2}- 1)$.  In the following $\Delta_{G_x}(\cdot)$ will always be with respect to $G_x$ and we drop the subscript.
For any $x$ it holds that $w_x(\Delta(V_0)) = |x|$ and $w_x(\Delta(\{v_0\})) = k$.
Now consider any $x$ and a subset $\emptyset \neq Y \subsetneq V$ different from $V_0$ or $V_1$.  We can prove that $w_x(\Delta(Y)) \ge k$.  To this end note that either 
$\emptyset \neq Y \cap V_0 \subsetneq V_0$ or $\emptyset \neq Y \cap V_1 \subsetneq V_1$.  First assume that the former is the case.  Then
\[
w_x(\Delta(Y))
= \sum_{u \in Y, v \notin Y} w_x(\{u,v\})
\ge \sum_{u \in Y \cap V_0, v \in V_0 \backslash Y} w_x(\{u,v\})
\ge k,
\]
as $k$ is the weight of a minimum cut in the complete weighted graph over $\floor{n/2}$ nodes with all edge weights $\tau$.  If instead $\emptyset \neq Y \cap V_1 \subsetneq V_1$ 
then a similar argument shows that $w_x(\Delta(Y)) \ge \tau (\ceil{n/2}-1) \ge k$.

Thus if $|x| < k$ then $\Delta(V_0)$ will be the unique minimum cut of $G_x$, and the weight of a minimum cut in $G_x$ will be $w_x(\Delta(V_0)) = |x|$. On the other hand, if 
$|x| > k$ then the weight of a minimum cut in $G_x$ will be $k$, which is realized by the star cut $\Delta(\{v_0\})$ (and potentially other cuts in $G_x$) but 
not by $\Delta(V_0)$ as $w_x(\Delta(V_0)) = |x| > k$.

Let $\Scal = \{x \in \{0,1\}^N : |x| \in \{k-1,k+1\} \}$ and $\Gcal = \{G_x : x \in \Scal\}$.  
Suppose there was a $T$ query algorithm in the adjacency matrix model that for any $G_x \in \Gcal$ with probability at least $2/3$ output the weight of a minimum cut in $G_x$.
If the output is $<k$ then we know that $|x| = k-1$ and if the output is $k$ then we know that $|x| = k+1$.  Moreover, any query to the adjacency matrix of $G_x$ can be simulated 
by a query to $x$, thus such an algorithm gives a $T$ query algorithm to determine if $|x| = k-1$ or $|x| = k+1$ when we are promised one of these is the case.  Since $\tau \le (\floor{n/2}-1)/2$ 
we have $2k+1 \le N$ and therefore we may apply Theorem \ref{thm:thresh} to obtain $T \in \Omega(\sqrt{N k}) = \Omega(n^{3/2}\sqrt{\tau})$.

Similarly, a $T$ query algorithm in the adjacency matrix model that for any $G_x \in \Gcal$ with probability at least $2/3$ outputs the shores of a cut realizing the minimum weight 
also implies a $T$ query algorithm to determine if $|x| = k-1$ or $|x| = k+1$.  In this case, if $|x| = k-1$ then the output must be $(V_0, \overline{V}_0)$ as these are the shores of the unique 
minimum cut in $G_x$.  On the other hand, if $|x| = k+1$ then $(V_0, \overline{V}_0)$ is not a correct output.  Thus the output of the algorithm lets us determine with probability at least $2/3$ if 
$|x| = k-1$ or $|x| = k+1$ and we again have $T \in \Omega(\sqrt{N k}) = \Omega(n^{3/2}\sqrt{\tau})$.
\end{proof}

\subsection{Adjacency array model}
Given adjacency array access to a graph with edge-weight ratio $\tau$, we showed an upper bound of $\tO(\sqrt{m n \tau})$ on the quantum query complexity of computing the weight of a minimum cut.
In this section we prove two distinct lower bounds, each of which is tight in a specific regime.
First we show that for any $\tau > 1$ there exists a family of dense graphs on $n$ vertices with edge-weight ratio $\tau$ for which computing the weight of a minimum cut requires $\Omega(n^{3/2})$ queries to the adjacency array.
This shows that the adjacency array upper bound of \cref{thm:main} is tight for dense weighted graphs with constant (but non-unit) edge-weight ratio.
Secondly and using a different approach, for any $1 \leq \tau \in O(n)$ we prove an $\Omega(\tau n)$ lower bound for a family of dense graphs with edge-weight ratio $\tau$.
This shows that we cannot get a quantum speedup when $\tau \in \Omega(n)$.

\subsubsection{Constant edge-weight ratio}

For the first bound we first need a claim about the minimum cuts of a complete weighted bipartite graph.
\begin{claim}
\label{clm:complete_bipartite}
Let $n \ge 8$ be a multiple of $4$ and $G=(L \sqcup R,w)$ be a weighted bipartite graph with bipartition $L,R$ where $|L| = 3n/4, |R| = n/4$. Further suppose that 
for every $x \in L, y \in R$ it holds that $w(\{x,y\}) \ge 1$.   Then any cut of $G$ that is not of the form $\Delta_G(\{x\})$ for $x \in L$ has weight at least $n/2$.  
\end{claim}

\begin{proof}
First consider a star cut $\Delta_G(\{y\})$ for $y \in R$.  This has weight at least $3n/4$, since this is the degree of $y$ and all edges have weight at least 1.  

It now remains to show the claim holds for non-star cuts.  Consider a general non-star cut with shore $X \cup Y$ with $X \subseteq L, Y \subseteq R$.  Let 
$k = |X|, \ell = |Y|$.  As it is a non-star cut we have $k + \ell \ge 2$.  By complementing as needed we may also assume that $k \le 3n/8$.  We also have 
the obvious constraints that $\ell \le n/4$ and $k,\ell \ge 0$.  

As $G$ is a complete weighted bipartite graph with every edge weight at least one we have
\[
w(\Delta_G(X \cup Y)) \ge k(n/4-\ell) + \ell(3n/4-k) \enspace .
\]
As $k \le 3n/8$ the term $\ell (3n/4-k)$ is greater than $n/2$ whenever $\ell \ge 2$.  Thus we can focus on $\ell \in \{0,1\}$.  
If $\ell=0$ then $k\ge 2$ and so the weight of the cut is at least $k(n/4)=n/2$ as desired.  If $\ell = 1$ then the weight of the 
cut is $k(n/4-1) + 3n/4-k$ which is always at least $3n/4$ as long as $n \ge 8$.  
\end{proof}

This claim means that if $\min_{x \in L} w(\Delta_G(\{x\})) < n/2$ then this value will be the weight of a minimum cut in $G$.  
We can leverage this to show a lower bound as follows.
In the next proof, for a function $f: \{0,1\}^n \rightarrow \{0,1\}$ we will use $Q_{1/3}(f)$ to denote the quantum query complexity of computing $f$ with error at most $1/3$.
\begin{theorem}
\label{thm:adjlower2}
Let $n \ge 8$ be a multiple of $4$ and $0 < \varepsilon \le 1$.  There is a family of $n$-vertex graphs $\Gcal$ all of which have edge weights in $\{1,1+\varepsilon\}$ 
such that any quantum algorithm that for every graph $G \in \Gcal$ computes with probability at least $2/3$ the weight of a minimum cut in $G$ must make 
$\Omega(n^{3/2})$ queries in the adjacency array model.  Similarly, any quantum algorithm that for every graph $G \in \Gcal$ computes with 
probability at least $2/3$ the shores $(X,\overline{X})$ of a cut realizing the minimum weight must make $\Omega(n^{3/2})$ queries in the adjacency array model.
\end{theorem}

\begin{proof}
Let $X =\{x \in \{0,1\}^{n/4}: |x| = \floor{n/8}-1\}$ and $Y =\{y \in \{0,1\}^{n/4}: |y| = \floor{n/8}+1\}$.  For every 
$x = (x^{(1)}, \ldots, x^{(3n/4)}) \in (X \cup Y)^{3n/4}$ we associate a bipartite graph $G_x = (L \sqcup R, w_x)$ 
where $L = \{1, \ldots, 3n/4\}, R = \{3n/4+1, \ldots, n\}$ and $w_x(\{i,j\}) = 1+ \varepsilon \cdot x^{(i)}(j-3n/4)$ for every $i \in L, j \in R$.  We set $\Gcal = \{G_x : x \in (X \cup Y)^{3n/4}\}$.

Define the function $g: X \cup Y \rightarrow \{0,1\}$ where $g(x) = 0$ iff $x \in X$.  We have $Q_{1/3}(g) \in \Omega(n)$ by \cref{thm:thresh}.  
Let $f : \{0,1\}^{3n/4} \rightarrow \{0,1\}$ be the AND function, for which $Q_{1/3}(f) \in \Omega(\sqrt{n})$.  By the composition theorem for quantum query complexity \cite{HLS07,Rei11a}, 
we have $Q_{1/3}(h) \in \Omega(n^{3/2})$ for the composed function $h = f \circ g^{3n/4}$.   

Let $x = (x^{(1)}, \ldots, x^{(3n/4)}) \in (X \cup Y)^{3n/4}$.  If $h(x) = 1$ then $x^{(i)} \in Y$ for all $i \in [3n/4]$ and the weight of the star cut $\Delta_{G_x}(\{i\}) = n/4 + \varepsilon \cdot(\floor{n/8}+1)$.  
As $\varepsilon \le 1$ this will be the weight of a minimum cut in $G_x$ by \cref{clm:complete_bipartite}.  On the other hand if $h(x) = 0$ then some $x^{(i)} \in X$ and 
$\Delta_{G_x}(\{i\}) = n/4 + \varepsilon \cdot(\floor{n/8}-1)$ and this will be the weight of a minimum cut of $G_x$.
Thus computing the weight of a minimum cut of $G_x$ lets us evaluate $h(x)$.  Further, given oracle access to $x$ we can 
simulate queries to $G_x$ in the adjacency array model.  Let $A_x$ be a $3n/4$-by-$n/4$ matrix whose $\ith$ row is the vector $1+ \varepsilon x^{(i)}$.  Then the vertical concatenation of $A$ with 
$A^T$ is a valid adjacency array for $G_x$.  To a degree query on vertex $i$ we simply answer $n/4$ if $1 \le 3n/4$ and $3n/4$ if $3n/4+1 \le i \le n$.  We can also answer 
a query to the name and weight of the $\jth$ neighbor of $i$ with one query to $x$.  This shows that the (1/3)-error quantum query complexity of computing the weight of 
a minimum cut on graphs in $\Gcal$ in the adjacency array model is at least $Q_{1/3}(f \circ g^{3n/4}) \in \Omega(n^{3/2})$.  

Finally, suppose a quantum query algorithm can compute a shore of a minimum cut in $G_x$ with $T$ queries.  We know that this shore must be of the form $\{v\}$ for a vertex $v \in L$.  
Thus with with $O(n)$ more queries the algorithm can classically compute the weight of a 
minimum cut by querying the weight of the neighbors of $v$.  Thus $T + O(n) \in \Omega(n^{3/2})$, which means $T \in \Omega(n^{3/2})$.  This completes the proof.
\end{proof}

\subsubsection{Large edge-weight ratio}
Let $n \in \N$ be a multiple of $4$ and $V$ be a vertex set with $|V| = n$.  Partition $V$ into four sets 
$V_1, V_2, V_3, V_4$.  

Now consider an integer $\tau$ such that $1 \leq \tau \leq 5n/8$ and $\tau n/10$ is an integer.  Fix a set $S$ of 
$\tau n/10$ ``edge disjoint'' quadruples $(u_1, u_2, u_3, u_4) \in V_1 \times V_2 \times V_3 \times V_4$.  By edge disjoint we mean no 
pair of consecutive elements $(u_i, u_{i+1})$ or $(u_4, u_1)$ appears in more than one quadruple. We fix an enumeration 
of $S$ and refer to the vertices in the $\ellth$ quadruple as $u_1^\ell, u_2^\ell, u_3^\ell, u_4^\ell$.  

\begin{figure}[htb]
\centering
\includegraphics[width=.7\textwidth]{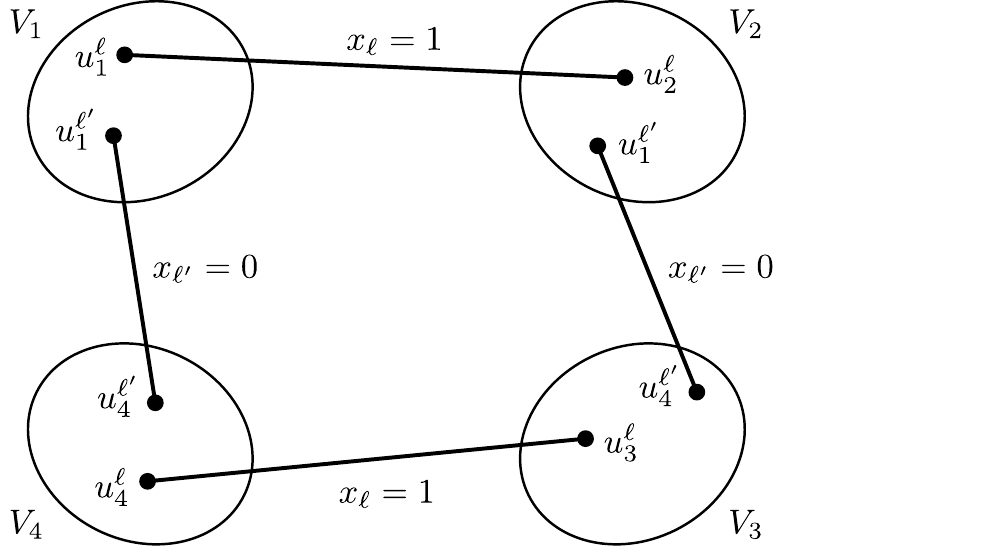}
\caption{Figure of graph $G_x$.
If $x_\ell = 1$ then we add edges $\{u_1^\ell,u_2^\ell\}$ and $\{u_3^\ell,u_4^\ell\}$.
If $x_\ell = 0$ then we add edges $\{u_2^\ell,u_3^\ell\}$ and $\{u_4^\ell,u_1^\ell\}$.}
\label{fig:adj-lower-bound}
\end{figure}

For every $x \in \{0,1\}^{\tau n/10}$ we define an $n$-vertex weighted graph $G_x = (V,w_x)$ where 
$w_x(\{u,v\}) = \tau$ if $u \ne v \in V_i$ for some $i \in [4]$, and for $\ell \in [\tau n/10]$ we set
\begin{align*}
w_x(\{u^\ell_1,u^\ell_2\}) &= w_x(\{u^\ell_3,u^\ell_4\}) = x_\ell, \\
w_x(\{u^\ell_2,u^\ell_3\}) &= w_x(\{u^\ell_4,u^\ell_1\}) = 1 - x_\ell \enspace.
\end{align*}
Otherwise, $w_x(\{u,v\}) = 0$.  
In words, on each $V_i$ we have a complete graph with all edge weights $\tau$, and for each $\ell \in [\tau n/10]$ we either add unit weight edges
$\{u^\ell_1,u^\ell_2\}, \{u^\ell_3,u^\ell_4\}$ or $\{u^\ell_2,u^\ell_3\},\{u^\ell_4,u^\ell_1\}$ depending on $x_\ell$.  
The construction is depicted in \cref{fig:adj-lower-bound}.

There are a few important points to note about this definition.  First, the edge-weight ratio of $G_x$ is $\tau$.  Second, for any $X$ that nontrivially intersects some $V_i$ we have that 
$w(\Delta_{G_x}(X)) \ge \tau(n/4-1)$.  This means that such an $X$ cannot be the shore of a minimum cut of $G_x$.  Third, by construction 
the degree of every vertex of $G_x$ is independent of $x$.  This means that degree queries to $G_x$ can be trivially answered and give us 
no information about $x$.  

\begin{lemma} \label{lem:adj-query}
We can simulate a single query to $G_x$ in the adjacency array model using a single query to $x$.
\end{lemma}
\begin{proof}
We first handle degree queries.  This can be answered with no queries to $x$ as the degree of a vertex is independent of $x$.

Now consider a query $(v,k) \in V \times [\deg(v)]$ to which we must answer the name $u$ of the $k$-th neighbor of 
$v$ and the edge weight $w_x(\{u,v\})$.  For clarity of exposition, we assume $v = u_1^t \in V_1$; the other cases are handled similarly.
\begin{itemize}
\item
If $k \leq n/4 - 1$ then return the $k$-th neighbor $u$ of $v$ inside $V_1$ and edge weight $w_x(\{u,v\}) = \tau$.
\item
If $k \ge n/4$ then let $j = k -n/4 +1$.  Letting $\ell$ denote the index of the $\jth$ quadruple of $S$ containing $v$ we query $x_\ell$.
\begin{itemize}
\item
If $x_\ell = 1$ then return neighbor $u^\ell_2$ and edge weight $w(\{v,u^\ell_2\}) = 1$.
\item
If $x_\ell = 0$ then return neighbor $u^\ell_4$ and edge weight $w(\{v,u^\ell_4\}) = 1$.
\end{itemize}
\end{itemize}
In total this takes a single query to $x$, which proves the lemma.
\end{proof}

\noindent
Now we can prove the following lemma.
\begin{lemma} \label{lem:adj-min-cut}
Fix integers $n$ and $\tau$ such that $1 \le \tau \le 5n/8$ and $\tau n/10 \in \N$.
Consider a string $x \in \{0,1\}^{\tau n/10}$ and the corresponding graph $G_x$.
If $|x| < \tau n/20$ then $G_x$ has a unique minimum cut with shores $(X,\overline X) = (V_1 \cup V_2, V_3 \cup V_4)$ and weight $w(\Delta_{G_x}(X)) = 2|x|$.
If $|x| > \tau n/20$ then $G_x$ has a unique minimum cut with shores $(X,\overline X) = (V_1 \cup V_4, V_2 \cup V_3)$ and weight $w(\Delta_{G_x}(X)) = 2(\tau n/10 - |x|)$.
\end{lemma}
\begin{proof}
First consider any cut shore that nontrivially intersects some $V_i$.
Since the subgraph $G_x[V_i]$ induced on $V_i$ is a complete graph with edge weights $\tau$, this implies that such a cut has weight at least $\tau (|V_i|-1) = \tau (n/4-1)$.
Now consider the small set of remaining cut shores that trivially intersect the $V_i$'s.
The weight of each one of these cuts can be easily expressed as a function of the Hamming weight $|x|$ of the input:
\begin{align*}
w(\Delta_{G_x}(V_i))
&= \tau n/10, \\
w(\Delta_{G_x}(V_1 \cup V_3)) &= w(\Delta_{G_x}(V_2 \cup V_4))
= 2 \tau n/10, \\
w(\Delta_{G_x}(V_1 \cup V_2)) &= w(\Delta_{G_x}(V_3 \cup V_4))
= 2 |x|, \\
w(\Delta_{G_x}(V_1 \cup V_4)) &= w(\Delta_{G_x}(V_2 \cup V_3))
= 2 (\tau n/10- |x|).
\end{align*}
It is clear that all minimum weight cuts will be among these cuts, and the lemma easily follows.
\end{proof}

Using this lemma we can prove the following theorem.

\begin{theorem}
\label{thm:adjlowerL}
Let $\tau, n \in \N$ be such that $1 \le \tau \le 5n/8$ and $\tau n/20 \in \N$.
There exists a family of $n$-vertex graphs $\Gcal'$ with $\Omega(n^2)$ edges, all of which have edge weights in $\{1,\tau\}$, such that any quantum algorithm that for every graph $G' \in \Gcal'$ computes with probability at least $2/3$ the weight of a minimum cut in $G'$ must make $\Omega(n \tau)$ queries in the adjacency array model.  Similarly, any quantum algorithm that for every graph $G' \in \Gcal'$ computes with probability at least $2/3$ the shores $(X,\overline{X})$ of a cut realizing the minimum weight must make $\Omega(n \tau)$ queries in the adjacency array model.
\end{theorem}
\begin{proof}
First consider the set of strings $\Xcal \subseteq \{0,1\}^{\tau n/10}$ with Hamming weight
\[
|x|
= \lfloor \tau n/100 \rceil \pm 1
< \tau n/20.
\]
By \cref{lem:adj-min-cut} the graph $G_x$, $x \in \Xcal$, has a unique minimum cut with shores $(V_1 \cup V_2, V_3 \cup V_4)$ and weight $2|x|$.
Now let $\Gcal' = \{G_x : x \in \Xcal\}$ and assume the existence of a quantum algorithm that for every $G_x \in \Gcal'$ computes with probability at least $2/3$ the weight $2|x|$ of a minimum cut in $G'$ with at most $q$ queries to the adjacency array of $G_x$.
By \cref{lem:adj-query} this is equivalent to outputting the Hamming weight $|x|$ with probability at least $2/3$ for any $x \in \Xcal$ while making only $q$ queries to $x$.
Using \cref{thm:thresh} this implies the lower bound $q \in \Omega(\tau n)$.

Next consider the set of strings $\Xcal' \subseteq \{0,1\}^{\tau n/10}$ that have Hamming weight $|x| = \tau n/20 \pm 1$.
By \cref{lem:adj-min-cut} the graph $G_x$, $x \in \Xcal'$, again has a unique minimum cut.
If $|x| = \tau n/20 - 1$ then its shores are $(V_1 \cup V_2, V_3 \cup V_4)$, while if $|x| = \tau n/20 + 1$ then its shores are $(V_1 \cup V_4, V_2 \cup V_3)$.
Now assume that there exists a quantum algorithm that with probability at least $2/3$ returns the shores of a minimum weight cut of $G_x$ with at most $q$ queries to the adjacency array of $G_x$.
By \cref{lem:adj-query} this is equivalent to distinguishing $|x| = \tau n/20 - 1$ from $|x| = \tau n/20 + 1$ with probability at least $2/3$ for any $x \in \Xcal$ while making only $q$ queries to $x$.
Using \cref{thm:thresh} this implies the lower bound $q \in \Omega(\tau n)$.
\end{proof}

\section*{Acknowledgements}
We would like to thank Ronald de Wolf for discussions which started this paper, and in particular a conversation which led to \cref{thm:lower}.
We also thank Debmalya Panigrahi and Miklos Santha for helpful conversations on this topic.
Simon Apers is supported in part by the Dutch Research Council (NWO) through QuantERA ERA-NET Cofund project QuantAlgo 680-91-034.
Troy Lee is supported in part by the Australian Research Council Grant No: DP200100950.


\newcommand{\etalchar}[1]{$^{#1}$}


\appendix
\section{Karger's theorem}
\label{app:karger}
In this appendix we prove a slight generalization of Karger's theorem \cite[Theorem 4.1]{Karger00} which is needed for our time-efficient algorithm.  
We begin by introducing some needed tools.

\subsection{Tools}
Matula \cite{Mat93} gave an $O(m/\varepsilon)$ time deterministic algorithm to compute a $(2+\varepsilon)$-approximation to the edge connectivity of a simple graph (or multigraph).  
The algorithm can also be adapted to give a constant factor approximation to the weight of a minimum cut in an integer-weighted graph in time $O(m \log^2(n))$, see Appendix~A of \cite{GMW20}.
\begin{lemma}[Matula's approximation algorithm \cite{Mat93, GMW20}]
\label{lem:matula}
Let $G = (V,w)$ be an integer-weighted graph with $m$ edges and $n$ vertices.  There is a constant $c$ and a deterministic algorithm that in time $O(m \log^2(n))$
outputs a value $\tilde \lambda$ such that $\tilde \lambda/c \le \lambda(G) \le  \lambda$.
\end{lemma}

To efficiently construct a tree-packing we will also need to use random sampling.  The following lemma is the heart of Karger's skeleton construction \cite{Karger99}.  We recommend 
the presentation in \cite[Lemma 14]{BLS20}.
\begin{lemma}[\cite{Karger99}]
\label{lem:downsample}
Let $G$ be an unweighted multigraph with $m$ edges.  For an integer $d \ge 2$ and real numbers $\varepsilon, \gamma$ with $\varepsilon \le 1/3$, 
let $p = 3d(\ln n)/(\varepsilon \lambda(G))$.  In time $O(pm \log(n))$  we can randomly sample $\ceil{pm}$ edges of $G$.  With probability $1-1/n^d$ the resulting graph $H$ has the properties that
\begin{enumerate}
  \item The minimum cut of $H$ is within a $(1+\epsilon)$ factor of $p\lambda(G) = 3d\ln(n)/\varepsilon^2$.
  \item For every $X \subseteq V$ we have $(1-\varepsilon) w(\Delta_G(X)) \le w(\Delta_H(X)) \le (1+\varepsilon) w(\Delta_G(X))$.
\end{enumerate}
\end{lemma}

Another very useful tool we use is the Nagamochi-Ibaraki construction which shows that for an integer-weighted graph $G$ with $m$ edges, in time $O(m \log(n))$ one can construct 
a graph $G'$ whose total edge weight is $nc$ and which preserves all cuts of $G$ of weight at most $c$.  

\begin{lemma}[\cite{NI92}]
Let $G =(V,w)$ be an $n$-vertex integer-weighted graph with $m$ edges.  For any positive integer $c$ there is a deterministic algorithm that in time $O(m \log n)$ produces an integer-weighted graph $G' = (V,w')$ with 
total edge weight $O(cn)$ such that for all $X \subseteq V$ with $\Delta_G(X) \le c$ it holds that $w(e) = w'(e)$ for all $e \in \Delta_G(X)$.  Thus in particular $\Delta_G(X) = \Delta_{G'}(X)$ and $w(\Delta_G(X)) = w'(\Delta_{G'}(X))$ for 
all $X$ with $\Delta_G(X) \le c$.
\end{lemma}

We combine the tools of Matula's approximation algorithm, random sampling, and the sparse certificate of Nagamochi-Ibaraki into the following lemma.
\begin{lemma}
\label{lem:Hprop}
Let $G = (V,w)$ be an integer-weighted graph and let $0 < \delta < 1$ be a parameter.  There is an $O(m \log^2(n) + n\log(n))$ time randomized algorithm to create a weighted graph $H = (V, w_H)$ such that 
\begin{enumerate}
  \item $H$ has $O(n \log(n)/\varepsilon^2)$ edges.
  \item The minimum cut of $H$ has value $\lambda(H) = O(\log n)$.
  \item If $X \subseteq V$ is such that $w(\Delta(X)) \le (1+\delta) \lambda(G)$ then $w_H(\Delta(X)) \le (1+3\delta) \lambda(H)$.
\end{enumerate}
\end{lemma}

\begin{proof}
First, by \cref{lem:matula}, in time $O(m \log^2(n))$ we can find a constant factor approximation $\tilde \lambda$ satisfying $\tilde \lambda/c \le \lambda(G) \le \tilde \lambda$ .  Next we apply the Nagamochi-Ibaraki 
algorithm to $G$ with threshold $t = (1+\delta)\tilde \lambda$.  In $O(m \log(n))$ time this produces an integer-weighted graph $G_2 = (V,w')$ with total edge weight $O(tn)$ such that for every $X \subseteq V$ with 
$w(\Delta_G(X)) \le (1+\delta) \tilde \lambda$ it holds that $w(\Delta_G(X)) = w'(\Delta_G(X))$.  

We now view $G_2$ as an unweighted multigraph with $O(tn)$ edges and apply \cref{lem:downsample}.
Let $p = \ln(n)/\tilde \lambda$.  We randomly choose $\ceil{p E(G_2)} =  O(n \ln(n))$ edges of $G_2$ and let the resulting graph be $H$.  This can be done in time $O(n \log(n))$.  
By \cref{lem:downsample} the graph has the stated properties.  The total running time is $O(m\log^2(n)+ n \log(n))$.
\end{proof}

\subsection{Tree packing}
With these preliminaries in place we now turn to actually constructing a tree packing.  We first need the definition, and a lemma of Karger.
\begin{definition}[Weighted tree packing]
Let $G = (V,w)$ be an integer-weighted graph.  A \emph{weighted tree packing} is a set of spanning trees of $G$, each with an assigned weight, such that 
the total weight of trees containing any edge $e \in E(G)$ is at most $w(e)$.  The \emph{value} of the packing is the total weight of trees in it.  
\end{definition}

\begin{lemma}[{\cite[Lemma 2.3]{Karger00}}]
\label{lem:frac_trees}
Given a weighted tree packing of value $\beta c$ and a cut of value $\alpha c$, at least a $(3-\alpha/\beta)/2$ fraction of the trees by weight 2-constrain the cut.
\end{lemma}

Gabow gives an algorithm to construct a near optimal tree packing in an unweighted multigraph.  The following is an easy adaptation to an integer-weighted graph.
\begin{lemma}[\cite{Gabow95}]
\label{lem:tree_packing}
Let $G = (V,w)$ be an integer-weighted graph with $n$ vertices and $m$ edges.  There is a deterministic algorithm that finds an integer-weighted tree packing of $G$ of value at least 
$\lambda(G)/2$ in time $O(m (\lambda(G)^2 \log(n) + \log^2(n)))$.  
\end{lemma}

\begin{proof}
For a multigraph $H$ with $n$ vertices and $m'$ edges, Gabow \cite{Gabow95} gives a deterministic algorithm that finds a tree packing of weight $\lambda(H)/2$ in time 
$m' \lambda(H) \log(n)$.  The only difference with our case is that $G$ is an integer-weighted graph instead of a multigraph.  We can of course view $G$ as a multigraph but it 
becomes too expensive to run Gabow's algorithm if this significantly blows up the number of edges.  

Thus we first use \cref{lem:matula} to compute $\tilde \lambda$ such that $\tilde \lambda/c \le \lambda(G) \le \tilde \lambda$ in time $O(m \log^2(n))$.  Then we make a pass through the 
edges of $G$ and form a graph $G'$ where any edge of weight larger than $\tilde \lambda$ in $G$ is thresholded down to $\tilde \lambda$.  Thus when viewed as a multigraph $G'$ will only have $O(m \lambda(G))$ edges.  Any tree packing of $G$ is also a tree packing of $G'$ as the value of any tree packing is at most $\lambda(G) \le \tilde \lambda$.  We can then apply Gabow's algorithm to $G'$ to obtain the theorem.
\end{proof}

We are finally ready to prove the slight generalization of Karger's theorem that we require.
\kargerthm*

\begin{proof}
In $O(m)$ time we can find the minimum weight $\alpha$ of an edge of $G$.  Multiplying all edge weights by $1/\alpha$ we obtain a graph where all edge weights are at least $1$ and that has the same 
set of $(1+1/16)$-near minimum cuts as $G$.  Thus without loss of generality now assume that $G$ has all edge weights at least $1$.  

In $O(m)$ time we create the integer-weighted graph $G' = (V,w')$ where $w'(e) = \lfloor 100w(e) \rceil$.  Note that as we assume that every edge of $G$ has weight at least $1$, for any 
$X \subseteq V$ we have
\begin{equation}
\label{eq:int_approx}
0.995 w(\Delta_G(X)) \le \frac{ w(\Delta_{G'}(X))}{100} \le 1.005 w(\Delta_G(X)) \enspace .
\end{equation}
Thus if $\Delta_G(X)$ is a $(1+ \varepsilon)$-near minimum cut of $G$ then $\Delta_{G'}(X)$ is a $(1+\varepsilon)(1.005)^2$-near minimum cut of $G'$.  With $\varepsilon = 1/16$ 
it follows that $\Delta_{G'}(X)$ is a $1+1/12$-near minimum cut of $G'$.

Next we apply \cref{lem:Hprop} to $G'$ to in time $O((m + n) \log^2(n))$ create a graph $H$ with the properties specified there.  We then use \cref{lem:tree_packing} to find a tree packing 
of weight at least $\lambda(H)/2$ and which contains $O(\log(n))$ trees since $\lambda(H) = O(\log(n))$.  Now let $\Delta_G(X)$ be a $(1+1/16)$-near minimum cut of $G$. 
Then $\Delta_{G'}(X)$ is a $(1+1/12)$-near mincut of $G'$ and by \cref{lem:Hprop}, $\Delta_H(X)$ is a $1+1/4$-near mincut of $H$.  
Therefore by \cref{lem:frac_trees} at least $1/4$ of the trees in the packing will $2$-respect $\Delta_H(X)$.  
These trees must also $2$-respect $\Delta_G(X)$ since it has the same shore $X$.
\end{proof}

\end{document}